\documentclass[aps,prc,twocolumn,superscriptaddress,floatfix,amsmath,amssymb,nofootinbib,longbibliography]{revtex4-2}

\usepackage{bm}
\usepackage{bbm}
\usepackage{dcolumn}
\usepackage{amsfonts}
\usepackage{amsthm}
\usepackage{bbold}
\usepackage{color}
\usepackage{braket}
\usepackage{graphicx}
\usepackage{multirow}
\usepackage{siunitx}
\usepackage[caption=false]{subfig}

\usepackage{orcidlink}
\usepackage{hyperref}
\hypersetup{
    colorlinks = true,
    citecolor  = blue,
    linkcolor  = blue,
    urlcolor  = blue
}

\newcommand{\eg}{e.g.}
\newcommand{\ie}{i.e.}
\newcommand{\ai}{\textit{ab initio}}

\newcommand{\magicint}{1.8/2.0 (EM)}
\newcommand{\deltago}{$\Delta$N$^2$LO$_\text{GO}$ (394)}
\newcommand{\emarthuis}{1.8/2.0 (EM7.5)}
\newcommand{\emsim}{1.8/2.0 (sim7.5)}
\newcommand{\nnlosat}{N$^2$LO$_\text{sat}$}

\newcommand{\fm}{\ensuremath{\text{fm}}}

\newcommand{\elem}[2]{\ensuremath{^{#2}\text{#1}}}

\begin{document}

\author{Z.~Li~\orcidlink{0000-0003-2786-7272}}
\email{zhen.li1@tu-darmstadt.de}%
\affiliation{Technische Universit\"at Darmstadt, Department of Physics, 64289 Darmstadt, Germany}
\affiliation{ExtreMe Matter Institute EMMI, GSI Helmholtzzentrum f\"ur Schwerionenforschung GmbH, 64291 Darmstadt, Germany}
\affiliation{Max-Planck-Institut f\"ur Kernphysik, Saupfercheckweg 1, 69117 Heidelberg, Germany}
\affiliation{LP2IB (CNRS/IN2P3 – Universit\'e de Bordeaux), 33170 Gradignan, France}

\author{A.~Tichai~\orcidlink{0000-0002-0618-0685}}
\email{alexander.tichai@tu-darmstadt.de}%
\affiliation{Technische Universit\"at Darmstadt, Department of Physics, 64289 Darmstadt, Germany}
\affiliation{ExtreMe Matter Institute EMMI, GSI Helmholtzzentrum f\"ur Schwerionenforschung GmbH, 64291 Darmstadt, Germany}
\affiliation{Max-Planck-Institut f\"ur Kernphysik, Saupfercheckweg 1, 69117 Heidelberg, Germany}

\author{A.~Schwenk~\orcidlink{0000-0001-8027-4076}}
\email{schwenk@physik.tu-darmstadt.de}%
\affiliation{Technische Universit\"at Darmstadt, Department of Physics, 64289 Darmstadt, Germany}
\affiliation{ExtreMe Matter Institute EMMI, GSI Helmholtzzentrum f\"ur Schwerionenforschung GmbH, 64291 Darmstadt, Germany}
\affiliation{Max-Planck-Institut f\"ur Kernphysik, Saupfercheckweg 1, 69117 Heidelberg, Germany}

\author{N.~A.~Smirnova~\orcidlink{0000-0001-8944-7631}}
\email{smirnova@lp2ib.in2p3.fr}%
\affiliation{LP2IB (CNRS/IN2P3 – Universit\'e de Bordeaux), 33170 Gradignan, France}

\allowdisplaybreaks

\title{High-order perturbative calculations of nuclear ground states: \\
Automated evaluation of many-body diagrams}

\begin{abstract}
We advance the many-body perturbation theory (MBPT) calculations of the ground-state energy and radius of closed-shell nuclei beyond third order. Using automated diagram generation and evaluation up to fifth order, we present ground-state properties of selected closed-shell nuclei up to $^{78}$Ni with two- and three-nucleon interactions derived from chiral effective field theory. A clear convergence trend is observed for the ground-state energy enabling calculations at improved accuracy. We further investigate in detail the decomposition of the fourth-order contributions. For the ground-state energy, the magnitude of the fourth-order contribution is typically less than half of the third order, and a typical cancellation among different classes of diagrams is observed. Finally, we perform a comprehensive comparison between MBPT and non-perturbative in-medium similarity renormalization group (IMSRG) calculations, with the goal to provide insight into many-body uncertainties associated with the IMSRG(2) truncation.
\end{abstract}

\maketitle

\section{Introduction}

\textit{Ab initio} calculations of atomic nuclei have achieved significant progress during the past decades~\cite{Hergert2020_review,Hebeler2021}. Starting from nucleon-nucleon (NN) and three-nucleon (3N) interactions based on chiral effective field theory (EFT)~\cite{Epelbaum2009,Machleidt2011}, the use of many-body frameworks with mild computational scaling has opened the door for the calculation of heavy nuclei~\cite{Hu2022ab,Hebeler2023,Miyagi:2023zvv,Tichai2023epe,Door:2024qqz,Bonaiti2025Pb,Kuske:2025tsm,Demol2026}. Examples of such methods are the in-medium similarity renormalization group (IMSRG)~\cite{Tsukiyama2011,Bogner2014,Hergert2016_IMSRG,Stroberg2019_review_Heff,Heinz2021}, coupled-cluster (CC) theory~\cite{Hagen2014,Novario2020a,Tichai2023epe,Marino2026}, self-consistent Green's function (SCGF) theory~\cite{Dickhoff2004,Soma2011,Barbieri2022,Brolli2025} and many-body perturbation theory (MBPT)~\cite{Tichai2016,Hu2016,Tichai2018,Tichai2020}. Improving the precision of many-body methods and quantifying the uncertainties are two of the key topics in the current advancement of \ai{} calculations. 

Many-body expansions are commonly approximated using two different truncation strategies: a truncation in terms of the operator rank and/or a truncation in terms of perturbative order. The former is commonly used in non-perturbative many-body methods, like IMSRG and CC, whereas the later is mainly used in MBPT. The operator-rank truncation in non-perturbative calculations allows one to efficiently resum a large class of diagrams to infinite order. For example, most IMSRG applications are truncated after normal-ordered two-body operators, whereas a more complete inclusion of three-body operators has been employed only in selected applications due to their high computational cost~\cite {Heinz2021,Heinz2025,He2024,Stroberg2024}. On the other hand, MBPT calculations for atomic nuclei are typically truncated at third order, referred to as MBPT(3), which can be efficiently evaluated at low computational cost throughout the entire nuclear chart~\cite{Tichai2020}. Therefore, advancing MBPT applications to fourth order and beyond not only increases the precision of nuclear calculations, but can also shed light on the theory uncertainties of MBPT as well as on non-perturbative many-body frameworks that only partially account for selected fourth- and fifth-order contributions in their respective expansions.

Extending MBPT calculations of finite nuclei beyond third order constitutes a major challenge due to the excessive number of diagrams that have to be derived and implemented. In addition, an efficient evaluation of MBPT diagrams commonly exploits rotational invariance of the nuclear Hamiltonian by casting the expressions into an angular-momentum-coupled form (the so-called $j$-scheme)~\cite{Kuo1981}. This derivation constitutes a major challenge in itself in particular once the number of many-body diagrams increases. To facilitate high-order calculations of atomic nuclei, we fully automate the workflow for diagram generation, angular-momentum coupling, and code generation. The initial steps of diagram generation have already been published~\cite{Arthuis2019ADG1,Arthuis2021ADG2,Tichai2022ADG3,Li2023thesis} and recently also in the context of automated Wick algebra for the IMSRG~\cite{Chen2026qcombo}. The MBPT calculation of nuclear matter has been advanced to third order for arbitrary proton fraction and temperature~\cite{Keller:2022crb,Alp:2025wjn}, to fourth order for neutron and symmetric matter~\cite{Drischler2019}, and recently to fifth order with automated diagram generation~\cite{Drischler2026}. In parallel, the automation of the error-prone angular-momentum coupling considerably simplifies the workflow in finite nuclei~\cite{Tichai2020AMC}. In this work, we develop the automation process for finite nuclei by generating a scalable MBPT code in \texttt{Fortran}, referred to as \texttt{MBPTFORT}. This enables MBPT calculations up to fifth-order for ground-state observables in closed-shell nuclei using modern chiral interactions. We finally emphasize that the use of automation techniques in quantum many-body theory has a long history in electronic structure calculations of chemical systems (see \eg{}, Refs.~\cite{Paldus1973,Janssen1991,Li1994,Evangelista2022}).

This paper is organized as follows. In Sec.~\ref{sec:theo_framework}, we briefly outline the framework of MBPT calculations of the ground-state energy and radius for closed-shell nuclei. We benchmark our automatically generated diagrams and code in Sec.~\ref{sec:medium_mass}, then we proceed with a convergence study in terms of perturbative orders and model-space size, and present the results of the calculation of ground-state energies and radii of selected closed-shell nuclei up to $^{78}$Ni. The sensitivity study with respect to nuclear interactions is also carried out in this section. We further decompose different diagrammatic contributions of the fourth-order correction and provide a detailed comparison between IMSRG(2) and MBPT calculations in Sec.~\ref{sec:comparison_imsrg2}. Finally, we conclude and give an outlook in Sec.~\ref{sec:summary}. 

\section{Theoretical framework}
\label{sec:theo_framework}

\begin{figure*}
\includegraphics[width=0.98\textwidth]{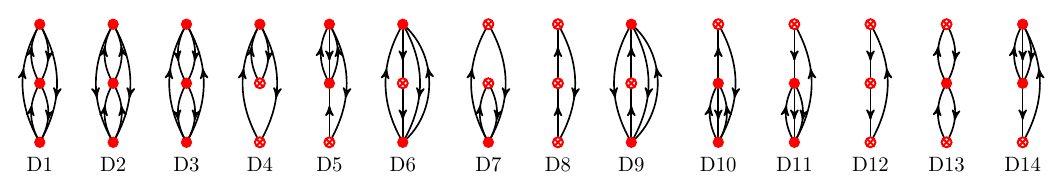}
\caption{\label{fig:thirdordermbpt}Third order Feynman diagrams in MBPT. Upward (downward) solid lines refer to particle (hole) lines. The one-body vertices $f_{ij}$ and two-body vertices $v_{ijkl}$ in the perturbed Hamiltonian $H_1$ are denoted by open circles with crosses and filled circles, respectively.}
\end{figure*}

\subsection{Reference state and partitioning}

In this section, we briefly outline the MBPT framework that is used for the calculation of ground-state energies and expectation values of scalar operators.
We start from a normal-ordered many-body Hamiltonian with zero-, one- and two-body terms 
\begin{eqnarray}
H = \mathcal{E}_0 + \sum_{ij} f_{ij} \{c_i^\dag c_j\} + \frac{1}{4} \sum_{ijkl} v_{ijkl} \{c_i^\dag c_j^\dag c_l c_k\} \,, 
\end{eqnarray}
where $i,j,k,l$ are single-particle states, $c^\dag$ ($c$) refers to the creation (annihilation) operator, and $f_{ij}$ and $v_{ijkl}$ are the normal-ordered one- and (anti-symmetrized) two-body matrix elements. The notation $\{\ldots\}$ denotes normal-ordering with respect to a many-body reference state 
\begin{align}
    |\Phi_0\rangle = \prod_{i=1}^A c^\dagger_i | 0 \rangle \, ,
\end{align}
that is taken as a spherical Hartree-Fock (HF) state in this work. Normal-ordered three- and higher-body terms are neglected here.
We partition the nuclear Hamiltonian according to
\begin{equation}
H = H_0 + H_1 \,,
\end{equation}
with the unperturbed Hamiltonian
\begin{eqnarray}
H_0 = \mathcal{E}_0 + \sum_{i} f_{ii} \{c_i^\dag c_i\} \, , 
\end{eqnarray}
and the residual perturbation
\begin{eqnarray}
H_1 = \sum_{i\neq j} f_{ij} \{c_i^\dag c_j\} + \frac{1}{4} \sum_{ijkl} v_{ijkl} \{c_i^\dag c_j^\dag c_l c_k\} \,. 
\end{eqnarray}
Computing the eigenvalues and eigenstates of $H_0$ is trivial
\begin{equation}
H_0|\Phi_m\rangle=\mathcal{E}_m|\Phi_m\rangle \,, \quad m = 0, 1, 2, \ldots, d-1 \,, 
\end{equation}
where the single-particle states are truncated so that the full model space is finite with dimension $d$, spanned by the unperturbed orthonormal many-body states $\{|\Phi_m\rangle,\,m=0,1,2,\ldots,d-1\}$ with $\mathcal{E}_0 \leq \mathcal{E}_1 \leq \cdots \leq \mathcal{E}_{d-1}$.

\subsection{Perturbative evaluation of ground-state energy}

The Schr\"odinger equation in the full model space can be projected into a smaller $d_p$-dimensional space $\mathbbm{P}\equiv\{|\Phi_m\rangle,\,m=0,1,2,\ldots,d_p-1\}$, and thus generating a $\mathbbm{P}$-space effective Hamiltonian~\cite{Suzuki1980} 
\begin{align}
    H_{\rm eff} = PH \Omega P \, ,
\end{align}
where $P \equiv \sum_{m=0}^{d_p-1} | \Phi_m \rangle \langle \Phi_m |$, and $\Omega$ denotes the wave operator that encodes all many-body correlations of the exact eigenstates in the full model space.
The wave operator satisfies the Bloch equation~\cite{Bloch1958,Suzuki1980}
\begin{eqnarray}
(\mathcal{E}-H_0) \Omega P = QH_1\Omega P - Q \Omega P PH_1\Omega P \, , 
\end{eqnarray}
for a degenerate $\mathbbm{P}$ space with $H_0P=\mathcal{E}P$, where $Q\equiv 1 - P$. 
The wave operator $\Omega$ can be expanded perturbatively as a power series
\begin{align}
\Omega  =  \sum_{n=0}^\infty \Omega^{(n)} \, ,
\end{align}
where the $(n+1)$-th-order correction $\Omega^{(n+1)}$ can be calculated iteratively using the recursion~\cite{Suzuki1980}
\begin{eqnarray}
\label{eq:omega_operator_recursive_equation}
\Omega^{(n+1)}
= R H_1 \Omega^{(n)} P - \sum_{k=1}^{n} R  \Omega^{(k)} P H_1 \Omega^{(n-k)} P \,,
\end{eqnarray}
starting from $\Omega^{(0)}=P$. In the above equation we have introduced the resolvent operator
\begin{align}
R \equiv Q\frac{1}{\mathcal{E}-QH_0Q}Q \, .
\end{align} 
For the special case of closed-shell nuclei, a simple one-dimensional $\mathbbm{P}$ space can be used, \ie{}, $P = | \Phi_0 \rangle \langle \Phi_0 |$ and $\mathcal{E}=\mathcal{E}_0$, and therefore the ground-state energy can be evaluated by  
\begin{align}\label{eq:energy_corr}
E = \langle\Phi_0 | H_{\rm eff} |\Phi_0 \rangle 
= \sum_{n=1}^{\infty} \langle \Phi_0 | H \Omega^{(n-1)} | \Phi_0 \rangle \, ,
\end{align}
with the $n$-th-order energy correction 
\begin{align} 
E^{(n)} \equiv \langle \Phi_0 | H \Omega^{(n-1)} | \Phi_0 \rangle \, .
\end{align}
Explicit expressions for the leading low orders are
\begin{subequations}    
\begin{align}
E^{(1)} &= \langle\Phi_0 | H |\Phi_0\rangle = \mathcal{E}_0 \,, \\
E^{(2)} &= \langle\Phi_0 | H_1 R H_1 | \Phi_0\rangle \,, \\
E^{(3)} &= \langle\Phi_0 | H_1 R H_1 R H_1 | \Phi_0\rangle \,, \\
E^{(4)} &= \langle\Phi_0 | H_1 R H_1 R H_1 R H_1 | \Phi_0\rangle \nonumber\\
&\quad - \langle\Phi_0 | H_1 R^2 H_1 | \Phi_0\rangle \langle\Phi_0 | H_1 R H_1 | \Phi_0\rangle \,, \\
&~\:\vdots \nonumber
\end{align}
\end{subequations}
where terms with $\langle\Phi_0 | H_1 |\Phi_0\rangle =0$ are not shown. Moreover, we denote the energy with corrections up to $n$th order as $E_{\rm sum}^{(n)} \equiv \sum_{k=1}^{n}E^{(k)}$.  The above formalism can be easily extended to open-shell system with a properly enlarged $\mathbbm{P}$ space to derive shell-model effective Hamiltonians~\cite{Coraggio2009,MHJ1995,Li2023thesis} and is equivalent to the formalism in, \eg{}, Refs.~\cite{Shavitt,Lietz2016}. 

Practically, the calculation of low-order energy corrections $E^{(n)}$ is efficiently organized using Feynman diagrams, see, \eg{}, Ref.~\cite{Shavitt}. Diagrams generated by the second term of Eq.~(\ref{eq:omega_operator_recursive_equation}) after being inserted into Eq.~(\ref{eq:energy_corr}) are all unlinked diagrams due to the middle $P$ in the expression, which cancel all unlinked diagrams generated by the first term, thus leading to a diagrammatic expansion containing only linked diagrams~\cite{Goldstone1957,Shavitt}.
As a consequence, the MBPT expansion is manifestly size-extensive and the energy corrections scale linearly with mass number.

For illustrative purpose, we present the Feynman diagrams in the third-order energy correction $E^{(3)}$ in Fig.~\ref{fig:thirdordermbpt}. The so-called canonical diagrams (D1-D3) including solely two-body $H_1$ vertices and the non-canonical diagrams with at least one one-body $H_1$ vertex (D4-D14) are shown. In the case of a canonical HF reference state, all non-canonical energy diagrams vanish by construction as the one-body Hamiltonian becomes diagonal, \ie{}, $f_{ij}=\delta_{ij} \, \epsilon_i$, where $\epsilon_i$ denotes the HF single-particle energy for the single-particle state $i$. 

\subsection{Operator expectation values}

The fully correlated ground state is restored by applying the wave operator on the reference state $|\Phi_0\rangle$ 
\begin{eqnarray}
|\Psi\rangle = \Omega |\Phi_0\rangle = \sum_{n=0}^{\infty} \Omega^{(n)} |\Phi_0\rangle \, .
\end{eqnarray}
This further allows us to calculate the expectation value of a scalar operator $\mathcal{O}$ in the ground state 
\begin{eqnarray}
\langle \mathcal{O}  \rangle
= \frac{\langle \Psi | \mathcal{O} | \Psi \rangle}{ \langle \Psi | \Psi \rangle} \, . 
\end{eqnarray}
This expectation value can also be conveniently calculated using Feynman diagrams by summing over all possible diagrams in which one $H_1$ vertex is replaced by one $\mathcal{O}$ vertex in the ground-state energy calculation~\cite{Brandow1967,Brandow1977,Miyagi2022_E3max}. In this way, a $n$-th-order ground-state energy diagram corresponds to $n$ different diagrams for the expectation value of $\mathcal{O}$. As an example, the third-order diagram D1 for the ground-state energy corresponds to three different diagrams for the expectation value of $\mathcal{O}$, 
\begin{equation}\label{eq:diag_ope_example}
\vcenter{\hbox{\includegraphics[width=0.4\textwidth]{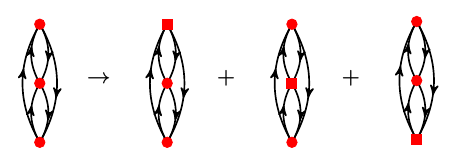}}}
\end{equation}
where the square vertex refers to the operator $\mathcal{O}$. In this work, we explicitly focus on the evaluation of the charge radius operator defined as
\begin{align}
R^2_\text{ch} \equiv R_p^2 + \langle r^2_\text{p} \rangle + \langle r^2_\text{n} \rangle \cdot \frac{N}{Z} + \langle r^2\rangle_\text{so} + r_\text{DF}^2 \, ,
\end{align}
where $R_p^2$ denotes the point-proton radius (from the center-of-mass), $\langle r_p^2 \rangle = 0.707\,\fm^2$ the proton charge radius squared~\cite{PDG2024}, $\langle r_n^2\rangle = -0.1155 \, \fm^2$ the neutron charge radius squared~\cite{PDG2024}, $\langle r^2\rangle_\text{so}$ the spin-orbit correction~\cite{Ong2010,Hoferichter:2020osn,Heinz2025} and $r_\text{DF}^2 = 3/(4M^2c^4) = 0.033 \, \fm^2$ the relativistic Darwin-Foldy term~\cite{Friar1997,Ong2010}. $N$ and $Z$ are the neutron and proton numbers, respectively. 

\subsection{Computer-aided implementation of Feynman diagrams}

The number of Feynman diagrams in MBPT contributing to the ground-state energy grows rapidly as the order increases. There are 2 (1), 14 (3), 201 (39), 4704 (840), 160890 (27300) total diagrams (diagrams with only two-body vertices) at the second, third, fourth, fifth, and sixth order, respectively. It is tedious and error-prone to find all the diagrams and derive their expressions by hand (especially for $j$-scheme expressions), beyond third order. In this work, we utilize the diagram generation and evaluation tool developed in Ref.~\cite{Li2023thesis} and the angular momentum coupling code \texttt{AMC}~\cite{Tichai2020AMC} to generate the $j$-scheme diagram expressions, as well as the corresponding \texttt{Fortran} code for numerical implementations with \texttt{MBPTFORT}. It is worth mentioning that there is another open-source Python package \texttt{ADG} developed by Arthuis \textit{et al.} (see Refs.~\cite{Arthuis2019ADG1,Arthuis2021ADG2,Tichai2022ADG3}), which is also able to generate and evaluate Feynman diagrams in MBPT and also in other many-body methods. By combining the public available softwares \texttt{ADG} and \texttt{AMC}, one is able to automate the diagram generation and evaluation in $j$-scheme in many-body calculations. 

\section{Many-body convergence in medium-mass nuclei}\label{sec:medium_mass}

\begin{figure}[t!]
\includegraphics[width=0.8\columnwidth]{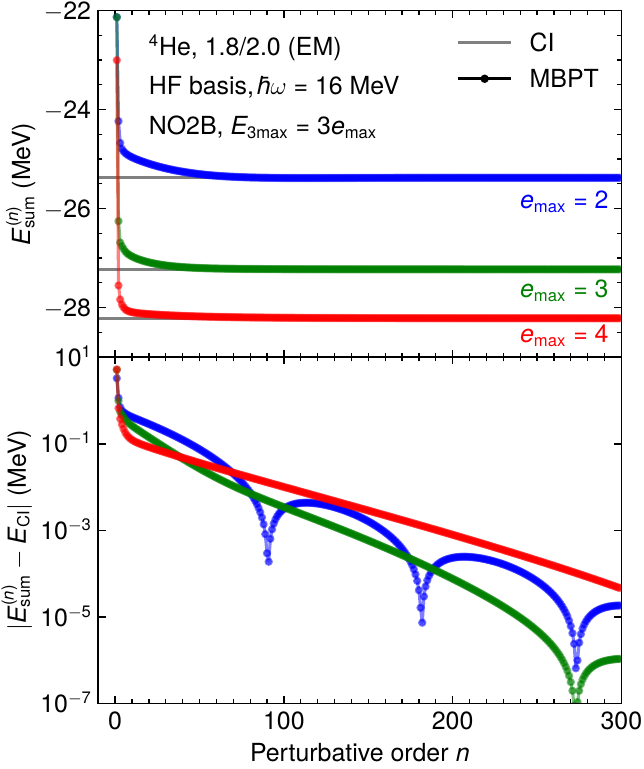}
\caption{MBPT ground-state energy (top) and its error relative to the exact CI ground-state energy (bottom panel) in small $e_{\rm max}$ model spaces as a function of perturbative order $n$ for $^{4}$He using the 1.8/2.0~(EM) interaction~\cite{Hebeler2011_EM1.8_2.0}. The Hamiltonian is truncated at the NO2B level in the HF basis. The recursive method is used in the MBPT calculation.}
\label{fig:recursive_calc}
\end{figure}

\subsection{Computational setup}

We start from a translationally invariant $A$-body Hamiltonian represented in the harmonic oscillator (HO) basis $H=T_{\rm rel} + V_{\rm NN} + V_{\rm 3N}$, consisting of relative kinetic energy $T_{\rm rel}$, NN interactions $V_{\rm NN}$, and 3N interactions $V_{\rm 3N}$. We initially perform a spherical HF calculation and transform the Hamiltonian into the HF basis. Furthermore, the Hamiltonian is normal-ordered with respect to the HF state, and we discard residual three-body terms giving rise to the so-called normal-ordered two-body (NO2B) approximation. The single-particle basis consists of 13 major oscillator shells, \ie{}, $e_{\rm max}=(2n+l)_{\rm max}=12$. The matrix elements of the 3N interaction $V_{\rm 3N}$ are further truncated by the three-body cut $E_{\rm 3max}=(e_1+e_2+e_3)_{\rm max}=24$~\cite{Miyagi2022_E3max}. Unless specified differently, we employ a HO frequency of $\hbar\omega=16$~MeV.  

\begin{figure}[t!]
\includegraphics[width=0.35\textwidth]{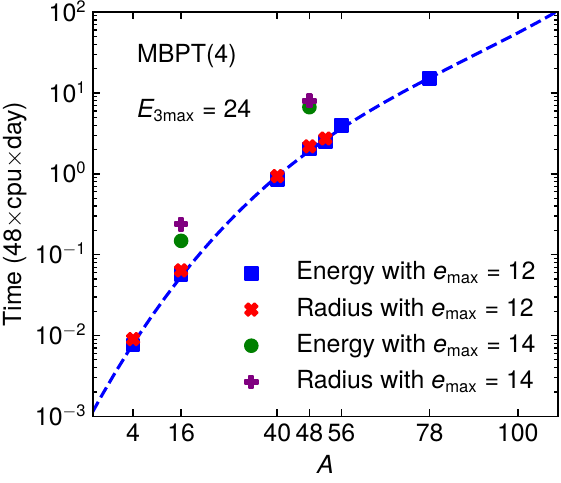} 
\caption{Required computing time as a function of mass number $A$ in the evaluation of the ground-state energy and charge radius up to the fourth-order correction (including solving the HF equation).}
\label{fig:time}
\end{figure}

In this work, different NN+3N chiral interactions are employed, the ``magic'' 1.8/2.0~(EM) interaction from Ref.~\cite{Hebeler2011_EM1.8_2.0}, the \deltago{} from Ref.~\cite{Jiang2020}, as well as the \nnlosat{} from Ref.~\cite{Jiang2020}.
We further study the recently developed \emarthuis{} and \emsim{} interactions from Ref.~\cite{Arthuis2024Hamil} that employ a variation of the $c_D$ (and correlated $c_E$) three-body coupling to improve the description of charge radii.

\subsection{Automated diagram evaluation: validation and performance}

\begin{figure*}
\includegraphics[width=0.7\textwidth]{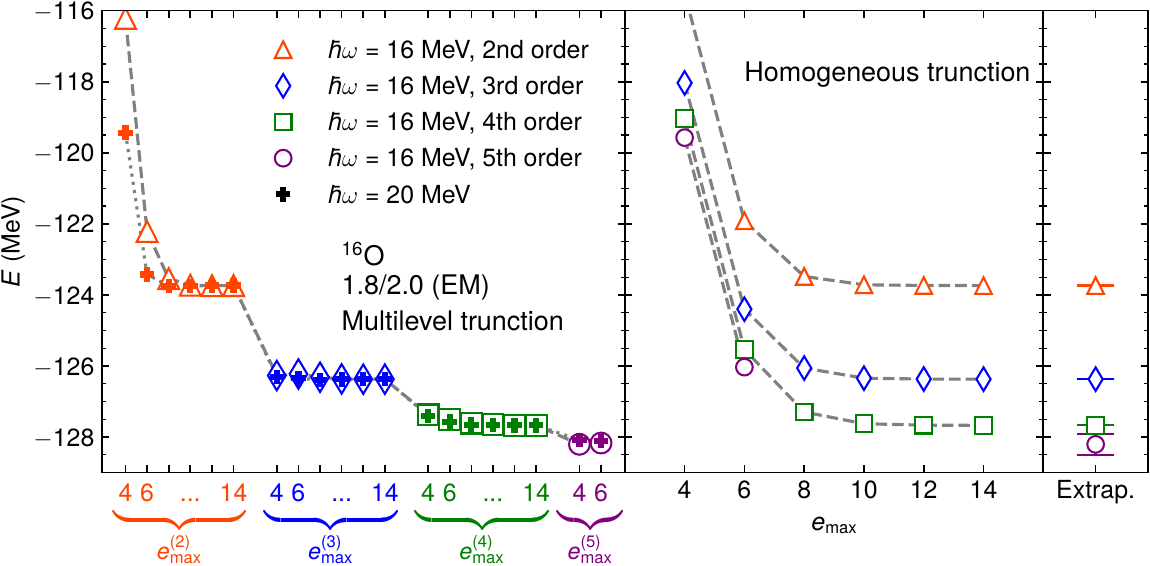} 
\caption{Convergence behavior with the multilevel truncation scheme (left) and the homogeneous truncation scheme (middle), as well as the extrapolation (right panel) for the MBPT ground-state energy of $^{16}$O calculated with the 1.8/2.0~(EM)~\cite{Hebeler2011_EM1.8_2.0} interaction. See text for details.}
\label{fig:multilevel_truncation}
\end{figure*}

We start our analysis by validating our computer-aided diagrammatic approach against a recursive implementation of MBPT energy corrections~\cite{Roth2010,Langhammer2012_Pade,Tichai2016}. To this end, we represent the operator from Eq.~(\ref{eq:omega_operator_recursive_equation}) in an $m$-scheme configuration basis and explicitly carry out high-order corrections up to order $n=300$. As the computational cost of such a recursive scheme is driven by the storage of many-body operators, it is equivalent in complexity to a no-core shell model (NCSM) calculation~\cite{Barrett2013}. Consequently, it cannot be extended to heavier systems and serves here as a benchmark to validate the correctness of our diagram/code generator. It is worth mentioning that a recent advancement within the framework of configuration-interaction quantum Monte Carlo~\cite{Zhen2026} may enable this recursive MBPT calculation to larger systems.

The ground-state energy of $^4$He obtained by our recursive implementation does reproduce the diagrammatic result up to machine precision in small benchmark model spaces with $e_{\rm max}=2,\,3,\,4$, thus ensuring correctness of our implementation. In addition, the MBPT expansion converges to the exact configuration-interaction (CI) result that is obtained by diagonalizing the many-body Hamiltonian in the same configuration basis using Lanczos techniques~\cite{Barrett2013}, as shown in Fig.~\ref{fig:recursive_calc}. We further observe that low-order results converge more rapidly in larger model space. These findings are consistent with previous work demonstrating that ``soft'' (low-resolution) chiral interactions are amenable to perturbative techniques~\cite{Tichai2016,Tichai2020}.

The automated generation of scalable implementations can be challenging and most production-level codes employ hand-optimized routines to speed up runtime-critical parts of their programs. We demonstrate that our implementation -- even though containing thousands of computer-generated diagrams -- leads to a scalable many-body framework applicable to nuclei with mass number of $A \lesssim 100$. Figure~\ref{fig:time} shows the consumed CPU time as a function of the mass number $A$ in the evaluation of the ground-state energy and nuclear charge radius up to the fourth-order MBPT. Most calculations are performed at $e_{\rm max}=12$ while for \elem{O}{16} and \elem{Ca}{48} calculations were further extended to $e_{\rm max}=14$. The computational cost of calculating the ground-state energy and radius can be made nearly identical, if the sum of the radius diagrams belonging to the same topology [\eg{}, the three radius diagram in Eq.~(\ref{eq:diag_ope_example})] is implemented in the innermost loop, although the radius calculation involves much more diagrams. We remark that the total complexity of the calculation is mainly driven by the diagrams with three-particle--three-hole excitations (triples) that appear at fourth order~\cite{Shavitt}, while the non-canonical diagrams with one-body vertices are zero for the ground-state energy and appear at minor computational cost for the radius due to fewer intermediate lines involved. 

\subsection{Model-space convergence}
\label{sec:multilevel_truncaton}

\begin{figure*}
\includegraphics[width=0.95\textwidth]{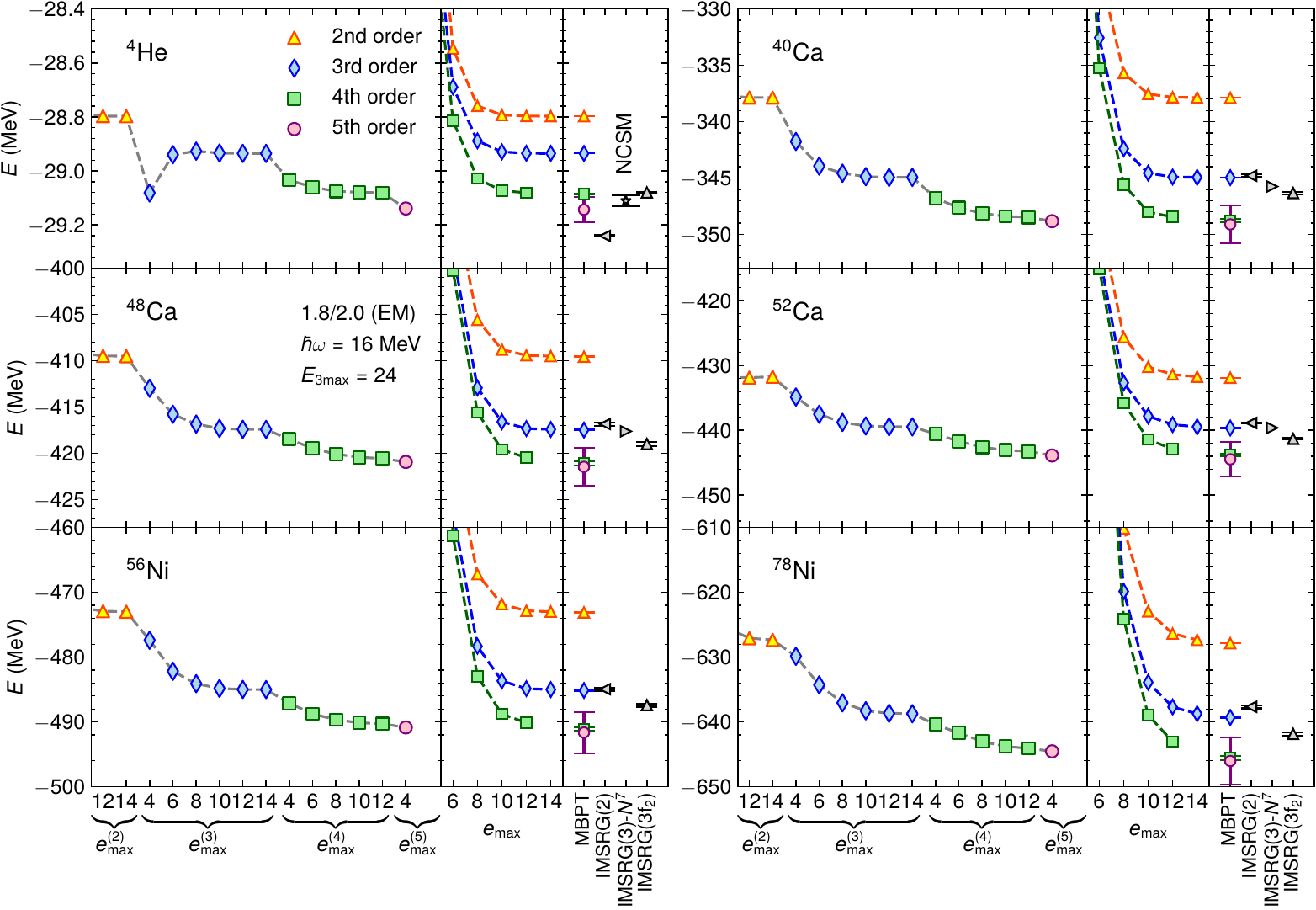}
\caption{MBPT results for the ground-state energies of selected closed-shell nuclei with the 1.8/2.0~(EM) interaction~\cite{Hebeler2011_EM1.8_2.0}. The convergence behavior with the multilevel truncation scheme and the homogeneous truncation scheme, as well as the MBPT extrapolation compared to IMSRG(2) and IMSRG(3f$_2$) with perturbative triples~\cite{He2024,Stroberg2024} are presented from left to right for each nucleus. The results from NCSM (for $^4$He)~\cite{Hebeler2021} and IMSRG(3)-$N^7$ (for the Ca isotopes)~\cite{Heinz2025} are also given. $e_{\rm max}=14$ is used in the HF calculation for the multilevel truncation. See text for details.}
\label{fig:binding_energy}
\end{figure*}

We continue with a detailed investigation of the model-space convergence pattern at different MBPT orders. To this end, we study two different truncation schemes: i) a multilevel truncation scheme that employs an order-specific truncation of the model space, and ii) a homogeneous truncation scheme where the same model-space truncation is used at all orders. As the computational cost at higher orders increases significantly, the multilevel truncation allows us to keep the overall cost tractable while including higher-order correlations in smaller spaces. Therefore, we introduce the parameter $e^{(n)}_{\rm max}$ to explicitly indicate the model-space size at order $n$.

The convergence behavior for the ground-state energy of $^{16}$O as a function of $e_{\rm max}^{(n)}$ is plotted in the left panel of Fig.~\ref{fig:multilevel_truncation} using the multilevel truncation, with the frequencies $\hbar\omega=16$~MeV (open symbols) and $\hbar\omega=20$~MeV (filled symbols). The model-space size is varied between $e_{\rm max}^{(n)} = 4$ and $14$ in all calculations up to the fourth order. We observe that in this multilevel truncation scheme, the third- and fourth-order contributions converge much faster compared to the second-order ones. This statement holds for various $\hbar \omega$ values. The calculation of the fifth-order correction is limited to $e^{(5)}_{\rm max}=6$ due to its high computational cost. For the employed values of $e_{\rm max}^{(5)} = 4,\,6$ the fifth-order corrections converge quickly with respect to $e_{\rm max}^{(5)}$, yielding a still non-negligible attractive correction to the energy of \elem{O}{16}.

In the middle panel of Fig.~\ref{fig:multilevel_truncation}, we present results for the homogeneous truncation scheme for $\hbar\omega=16$~MeV, where the same $e_{\rm max}$ is employed throughout all orders. The strong $e_{\rm max}$ dependence of the results up to third, fourth, and fifth order [referred to as MBPT(3), MBPT(4), and MBPT(5), respectively] is mainly due to the $e_{\rm max}$ dependence of the reference-state energy $\mathcal{E}_0$ and of the leading second-order correction $E^{(2)}$. Therefore, the MBPT(5) calculation in a small model space, e.g., $e_{\rm max}=6$, does not yield meaningful results within this homogeneous truncation scheme, as $\mathcal{E}_0$ and $E^{(2)}$ are far from converged.

The results of an extrapolation performed for $\hbar\omega=16$~MeV are presented in the right panel of the Fig.~\ref{fig:multilevel_truncation}. The extrapolation uses the method of Ref.~\cite{Furnstahl2012} and relies on the ground state energies obtained at MBPT(2), MBPT(3), and MBPT(4) within the homogeneous truncation scheme with $e_{\rm max}$ being varied from 8 to $14$. The uncertainty associated with each order is due to the statistical uncertainty during the fitting procedure. 
The extrapolated fifth-order value, referred to as MBPT(5*) (asterisk stands for multilevel truncation scheme is employed), is obtained by adding the fifth-order correction $E^{(5)}$ obtained with $e_{\rm max}^{(5)}=6$ on top of the extrapolated fourth-order result MBPT(4), with an additional uncertainty arising from the not yet fully converged $E^{(5)}$ in terms of $e_{\rm max}^{(5)}$. This uncertainty is taken from the energy variation of the fourth-order correction $E^{(4)}$ from $e_{\rm max}^{(4)}=4$ to $14$ to probe the effect due to the model-space truncation in the fifth-order calculation.

This multilevel truncation scheme can also be applied to radius calculations. Unlike the ground-state energy, the convergence of the radius calculation is less trivial and depends on the nucleus. We leave this discussion to Sec.~\ref{sec:radius}. 

\subsection{Ground-state energies of medium-mass nuclei}\label{sec:binding_energy}

\begin{figure*}[t!]
\includegraphics[width=0.95\textwidth]{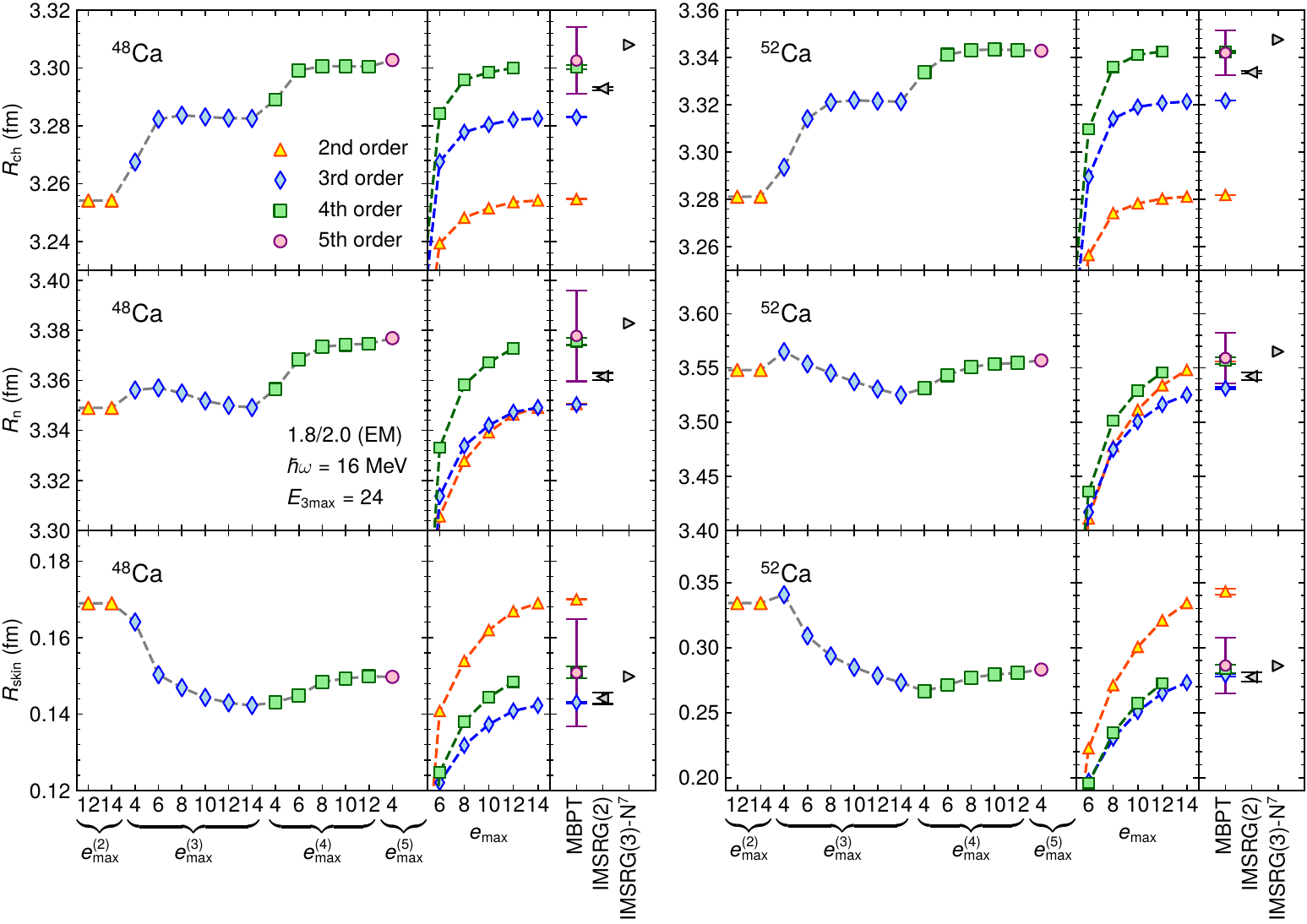} 
\caption{MBPT results for charge radii (top), neutron radii (middle), and neutron skins (bottom panels) for $^{48}$Ca and $^{52}$Ca with the 1.8/2.0~(EM) interaction~\cite{Hebeler2011_EM1.8_2.0}. The convergence behavior is shown for the multilevel truncation scheme and the homogeneous truncation scheme, and the MBPT extrapolations are compared to IMSRG(2) and IMSRG(3)-$N^7$~\cite{Heinz2025}. The HF calculation in the multilevel truncation employs $e_{\rm max}=14$. The updated value, $\langle r_p^2 \rangle = 0.707\,\fm^2$~\cite{PDG2024}, is used for all charge-radius results, including IMSRG(3)-$N^7$, where additional corrections have been made~\cite{Heinz2025}. See text for details.}
\label{fig:nuclear_radius}
\end{figure*}

Based on the detailed analysis for \elem{O}{16}, we extend our MBPT calculations of ground-state energies in Fig.~\ref{fig:binding_energy} to selected closed-shell nuclei from \elem{He}{4} to \elem{Ni}{78} using the \magicint{} interaction~\cite{Hebeler2011_EM1.8_2.0}. Our MBPT calculations are compared to non-perturbative IMSRG(2) and IMSRG(3f$_2$) (with perturbative triples)~\cite{He2024,Stroberg2024} calculations in all cases and additional IMSRG(3)-$N^7$~\cite{Heinz2021,Heinz2025} calculations for the Ca isotopes. For the light nucleus \elem{He}{4} we furthermore include the NCSM result from Ref.~\cite{Hebeler2021} as a many-body benchmark.

We observe in all cases that the fourth-order contribution converges much faster compared to the third order with respect to $e_{\rm max}$, as already discussed in Sec.~\ref{sec:multilevel_truncaton}. Based on these findings, we expect that fifth-order calculations converge equally quickly as a function of model-space size. As a result, the multilevel truncation with $e_{\rm max}^{(1,2,3)}=14$, $e_{\rm max}^{(4)}=12$, and $e_{\rm max}^{(5)}=4$ is a reasonable choice to obtain reliable results for the ground-state energy. However, a robust model-space extrapolation within the multilevel truncation scheme is challenging. We therefore stick to the homogeneous truncation scheme up to the fourth order and carry out model-space extrapolation~\cite{Furnstahl2012} using the results from $e_{\rm max}=6$ to $12$ to obtain MBPT(2,3,4), while applying the multilevel truncation scheme only to the fifth-order result, \ie{}, adding the $e_{\rm max}^{(5)}=4$ fifth-order contribution $E^{(5)}$ on top of the extrapolated fourth order result to give MBPT(5*). For a meaningful comparison, the IMSRG(2) and IMSRG(3f$_2$) results are the extrapolated values using the results from $e_{\rm max}=6$ to $12$. We see that the IMSRG(2) results are typically close to MBPT(3) results; a more detailed comparison between MBPT and IMSRG will be given in Sec.~\ref{sec:mbpt_vs_imsrg2}. The IMSRG(3)-$N^7$ results from Ref.~\cite{Heinz2025} (with the natural orbital basis and a different model space truncation strategy) lie between MBPT(3) and MBPT(4). The same is true for the IMSRG(3f$_2$) results, although their values are slightly lower than the IMSRG(3)-$N^7$ results for the Ca isotopes. 

Figure~\ref{fig:binding_energy} also reveals a clear monotonic convergence (we use the term ``convergence'' to refer to improvement in accuracy as perturbative order increases) trend in terms of perturbative orders with the \magicint{} interaction, \ie{}, higher-order calculations always yield lower energies. The size of the fourth-order correction $E^{(4)}$ is in general less than half of the third order one. The fifth-order correction $E^{(5)}$ is also smaller in absolute value than the fourth order one and it further lowers the ground-state energy, indicating that MBPT(5*) provides a better accuracy. In Sec.~\ref{sec:interaction_sensitivity}, we examine the sensitivity of our results to the choice of the interaction.

\subsection{Nuclear radii}
\label{sec:radius}

Following the same strategy as in Sec.~\ref{sec:binding_energy}, we carry out MBPT calculations of nuclear radii for $^{48}$Ca and $^{52}$Ca. As seen from Fig.~\ref{fig:nuclear_radius} (top panels), the charge radii exhibit a clear monotonic convergence trend, in both isotopes leading to larger radii value as the perturbative order increases. The fourth-order correction is smaller in magnitude than the third order one, but still sizable for both nuclei. Furthermore, we observe that the fourth-order result converges faster than the third-order one with respect to $e_{\rm max}$, similar to the convergence pattern of the ground-state energy. 

The convergence behavior of neutron radii $R_n$ and, therefore, neutron skins defined as $R_{\rm skin}=R_n-R_p$, is qualitatively different as can be seen from the middle and bottom panels of Fig.~\ref{fig:nuclear_radius}. Both quantities converge slower with respect to the basis size, especially in the case of $^{52}$Ca, which has a larger neutron excess. Therefore, the uncertainty associated to the fifth-order result may be unreliable for neutron radii and neutron skins. In addition, there is no clear order-by-order convergence trend, at least up to the fourth/fifth order. For example, the third-order correction to the neutron radius is small and positive in \elem{Ca}{48}, but is negative in \elem{Ca}{52}. In both cases, the fourth-order correction is larger in absolute value than the third-order one. Nevertheless, the uncertainty associated with the perturbative truncation is likely to be still smaller than the Hamiltonian uncertainty, which will be discussed in Sec.~\ref{sec:interaction_sensitivity}. 

We finally compare the MBPT results with non-perturbative IMSRG calculations performed at the IMSRG(2) and IMSRG(3)-$N^7$ level~\cite{Heinz2025}. The charge radii obtained within MBPT(2) substantially deviate from IMSRG(2) results, indicating the importance of third- and fourth-order contributions. The final MBPT(4) radii reside between the IMSRG(2) and IMSRG(3)-$N^7$ values, typically a bit closer to the more refined IMSRG(3)-$N^7$ predictions. As for neutron radius and neutron skin, the fourth-order results lie in the vicinity of the IMSRG(3)-$N^7$ values, although our calculations are not fully converged yet. 

\subsection{Interaction sensitivity}
\label{sec:interaction_sensitivity}

\begin{figure*}[t!]
\includegraphics[width=0.75\textwidth]{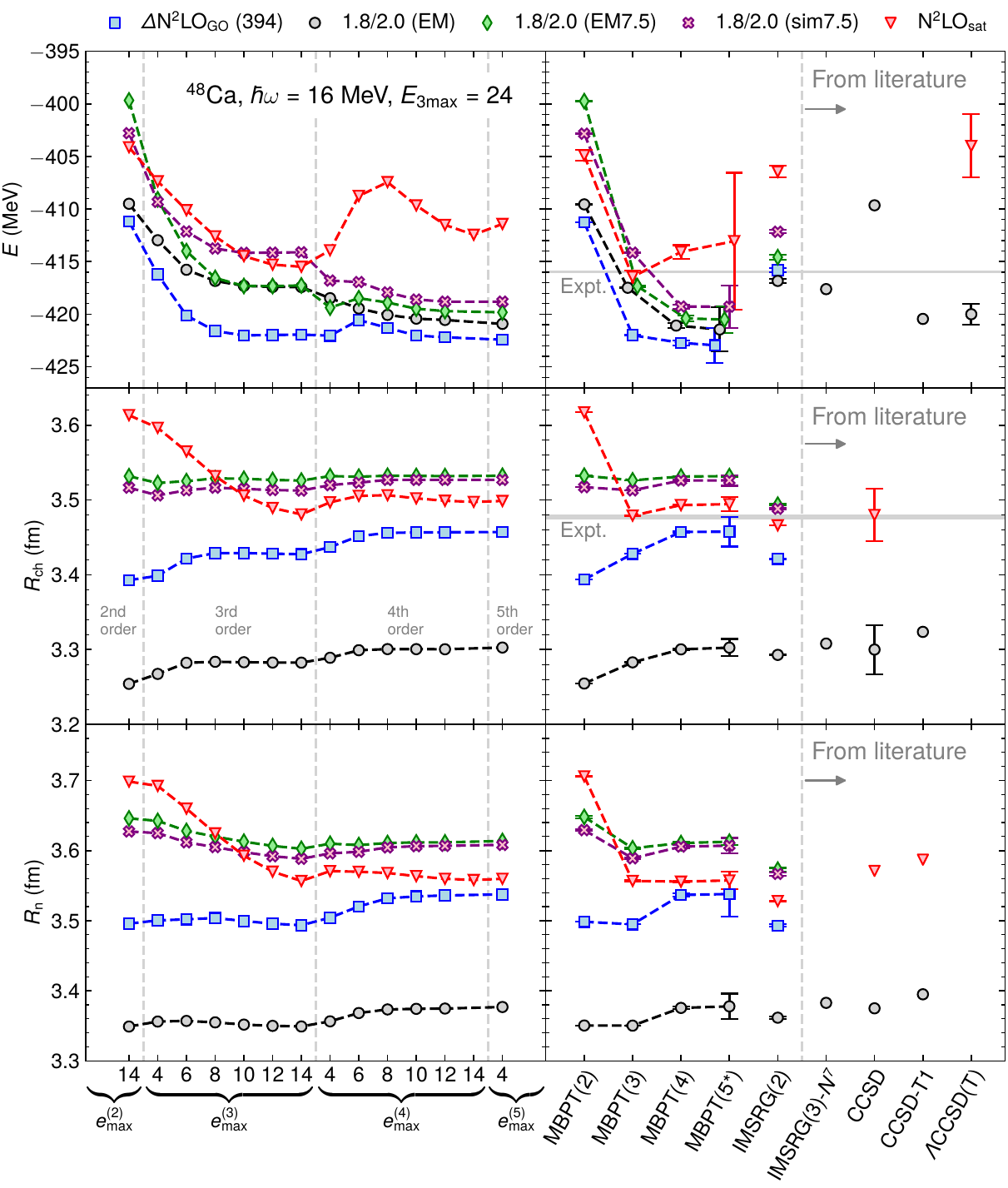}
\caption{MBPT results for the ground-state energy (top), charge radius (middle), and neutron radius of $^{48}$Ca (bottom panels) using the \deltago{}~\cite{Jiang2020}, \magicint{}~\cite{Hebeler2011_EM1.8_2.0}, \emarthuis{}~\cite{Arthuis2024Hamil}, \emsim{}~\cite{Arthuis2024Hamil}, and \nnlosat{}~\cite{Ekstrom2015_NNLOsat} interactions, compared to IMSRG(2) results in the same basis space and same extrapolation procedure. The MBPT(2,3,4) results for all interactions are obtained from model-space extrapolations~\cite{Furnstahl2012} in the homogeneous truncation scheme with $e_{\rm max}=6,\,8,\,10,\,12$ ($e_{\rm max}=8,\,10,\,12,\,14$ for \nnlosat{} due to its slow convergence), while MBPT(5*) refers to the multilevel truncation result obtained by adding the fifth-order correction calculated with $e_{\rm max}^{(5)}=4$ on top of MBPT(4). The uncertainties of MBPT(2,3,4) and IMSRG(2) come from the extrapolation, while an additional component due to the lack of convergence with respect to $e_{\rm max}^{(5)}$ is included in MBPT(5*) (see Sec.~\ref{sec:multilevel_truncaton}). Where available, literature results are shown for IMSRG(3)-$N^7$~\cite{Heinz2025}, CC with single and double excitations (CCSD)~\cite{Hagen2015Ca48,BonaitiCCresults,Simonis2019}~[top panel: $\hbar\omega=16$~MeV, $e_{\rm max}=14$, $E_{\rm 3max}=28$; middle panel: $\hbar\omega=22$~MeV, $e_{\rm max}=14$, $E_{\rm 3max}=18$ for \nnlosat{} and $E_{\rm 3max}=16$ for \magicint{}; bottom panel: $\hbar\omega=16$~MeV, $e_{\rm max}=14$, $E_{\rm 3max}=16$], CCSD-T1~\cite{Simonis2019,BonaitiCCresults}~($\hbar\omega=16$~MeV, $e_{\rm max}=14$, $E_{\rm 3max}=16$), and $\Lambda$-CCSD(T)~\cite{Hagen2015Ca48} [$\hbar\omega=22$~MeV, $e_{\rm max}=14$, $E_{\rm 3max}=16$--$18$ for \nnlosat{} and $E_{\rm 3max}=14$--$16$ for \magicint{}].}
\label{fig:Ca48_with_different_int}
\end{figure*}

\begin{figure}[t!]
\includegraphics[width=0.35\textwidth]{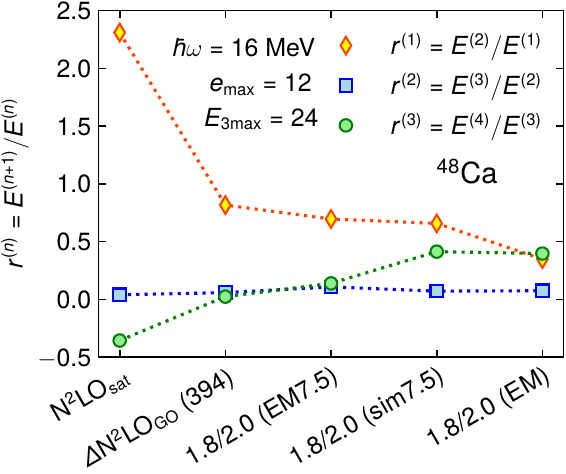} 
\caption{Ratio of successive perturbative contributions to the ground-state energy of $^{48}$Ca with different interactions.}
\label{fig:ratio_ca48_diff_int}
\end{figure}

The performance and accuracy of MBPT calculations are highly sensitive to the choice of the underlying inter-nucleon interactions. 
In this section, we study the impact of the Hamiltonian on the ground-state energy and charge/neutron radius for the case of $^{48}$Ca as a representative medium-mass nucleus. For this purpose, in Fig.~\ref{fig:Ca48_with_different_int} we compare the results obtained from four different ``softer'' interactions: \deltago{} \cite{Jiang2020}, \magicint{} \cite{Hebeler2011_EM1.8_2.0}, \emarthuis{} \cite{Arthuis2024Hamil}, and \emsim{} \cite{Arthuis2024Hamil}, as well as from the ``harder'' \nnlosat~interaction~\cite{Ekstrom2015_NNLOsat}.

For the ground-state energy, we observe a similar convergence behavior among the different ``softer'' interactions: 
Higher-order contributions converge faster as a function of $e_{\rm max}$, remain attractive, and decrease in magnitude as the MBPT order increases. The third-, fourth-, and fifth-order results obtained from all interactions consistently lie below the corresponding IMSRG(2) results, while the second-order results are all higher than IMSRG(2). More precisely, the IMSRG(2) results agree more closely with the MBPT(3) predictions, except for the \nnlosat{} and \deltago{} interactions. For the \magicint{} interaction, the ground-state energy from the more refined IMSRG(3)-$N^7$ calculation is about 1\,MeV lower than IMSRG(2), whereas the results from CCSDT-1/$\Lambda$-CCSD(T) are approximately 3-4\,MeV lower, providing better agreement with the MBPT(4)/MBPT(5*) results. We highlight, that the ground-state energy exhibits a clear monotonic convergence trend up to fifth-order with the set of ``softer'' interactions, especially with the \deltago~interaction, although full convergence is not yet certain.
The convergence pattern changes notably for the ``harder'' interaction \nnlosat{}. We observe an oscillatory behavior from one perturbative order to the next, which casts doubt on the convergence of the perturbative expansion. Consequently, the extrapolated energy carries a significantly larger uncertainty. Nevertheless, the fourth-order correction shifts the result in the direction toward the non-perturbative CCSDT-1/$\Lambda$-CCSD(T) results, which include all MBPT(4) diagrams.

It is worth noting that some fourth-order diagrams are missing in the non-perturbative IMSRG(2) calculation. We will compare  MBPT to IMSRG(2) in detail in Sec.~\ref{sec:mbpt_vs_imsrg2}. Although IMSRG(3)-$N^7$ is also fourth-order MBPT complete, the difference between IMSRG(3)-$N^7$ and MBPT(4) is quite large compared to the difference between CCSDT-1/$\Lambda$-CCSD(T) and MBPT(4), which may be attributed to the inclusion of additional commutators scaling up to $N^7$ (with $N$ being the number of single-particle states) in IMSRG(3)-$N^7$~\cite{Heinz2025}.

The results for the charge radius and neutron radius indicate that the uncertainties due to the perturbative truncation are much smaller than those associated with the choice of the interaction. However, as was noted above, the convergence properties remain  unclear up to fifth order. As expected~\cite{Simonis:2017dny}, the \magicint{} interaction produces a smaller charge radius and neutron radius even with the fourth- and fifth-order corrections, compared to the other interactions.

To provide further insight into the quantification of many-body uncertainties, we follow Ref.~\cite{Svensson2026} and investigate the ratio of successive energy corrections
\begin{align}
    r^{(n)} = \frac{E^{(n+1)}}{E^{(n)}}\, .
\end{align}
A ratio of $|r|<1$ indicates a convergent many-body expansion under the assumption of constant geometric suppression. Moreover, the ratio $r^{(n)}$ reflects the rate of convergence and is linked to the softness of the interaction.

We display the ratio $r^{(n)}$ up to $n=3$ for the five chiral interactions in Fig.~\ref{fig:ratio_ca48_diff_int}. The results are again for \elem{Ca}{48} as a representative medium-mass nucleus, and the ratios are expected to be approximately nucleus-independent by virtue of size-extensivity. The MBPT ratios exhibit a clear dependence on the choice of interaction, with the case of \nnlosat{} showing a particularly different trend. Indeed, the leading ratio $r^{(1)}$ is very large. This is due to the poorer HF approximation for this interaction, which captures only about $30\%$ of the total ground-state energy. The other, ``softer'', interactions yield much smaller $r^{(1)}$ ratios throughout. Interestingly, the subleading ratio $r^{(2)}$ is similar for all interactions and is very small in magnitude. This indicates that the third-order energy correction is strongly suppressed compared to the second-order one -- typically by a factor 10 or more. Based on the third-order energy corrections one might be overly confident in the rate of convergence. However, systematic fourth-order calculations presented in this work reveal that this suppression observed at third order may be accidental. In particular, for the \magicint{} interaction, the fourth-order ratio $r^{(3)}$ increases to $0.4$, calling for more conservative error models. Note that $r^{(3)}$ is similar for \emsim{}, but not for the \deltago{} and \emarthuis{} interactions, pointing towards a non-trivial interaction dependence. Although the origin of this feature is not yet fully understood, the enhanced fourth-order contribution may stem from the qualitative differences in their respective diagrams: higher excitations enter at fourth order for the first time. This is different to the decrease of the ratio from second to third order, where both orders probe the same excitation rank, \ie{}, two-particle--two-hole configurations.
In summary, the fourth-order energy corrections indicate that the assumption of a uniform geometric suppression model for uncertainty quantification has to be revisited for some interactions.

\section{MBPT versus IMSRG(2)}
\label{sec:comparison_imsrg2}

To examine the reliability of the MBPT calculation and gain insight in the uncertainty associated with the IMSRG(2) truncation, we make a detailed comparison of the non-perturbative content of the IMSRG(2) with perturbative calculations. 

\subsection{Anatomy of fourth-order contributions}

\begin{figure}[t!]
\includegraphics[width=0.4\textwidth]{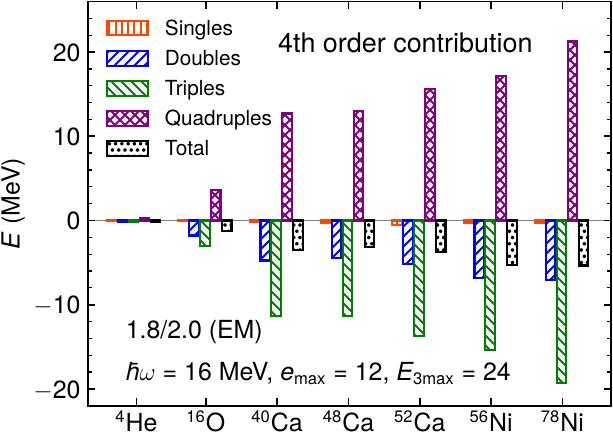}  
\caption{Contributions from singles, doubles, triples, and quadruples to the fourth-order correction of the ground-state energy of selected closed-shell nuclei for the \magicint~interaction~\cite{Hebeler2011_EM1.8_2.0}.}
\label{fig:o4_all_nuclei}
\end{figure}

\begin{figure}[t!]
\includegraphics[height=0.335\textwidth]{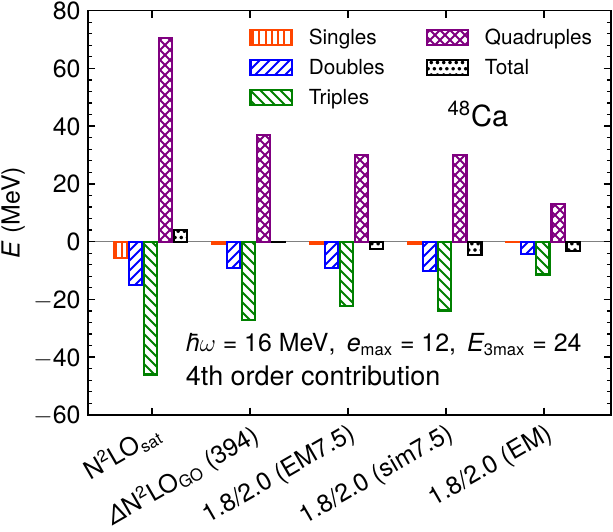}  
\caption{Contributions from singles, doubles, triples, and quadruples to the fourth-order correction of the ground-state energy of $^{48}$Ca for different interactions.}
\label{fig:e4_all_int}
\end{figure}

\begin{figure}[t!]
\includegraphics[height=0.335\textwidth]{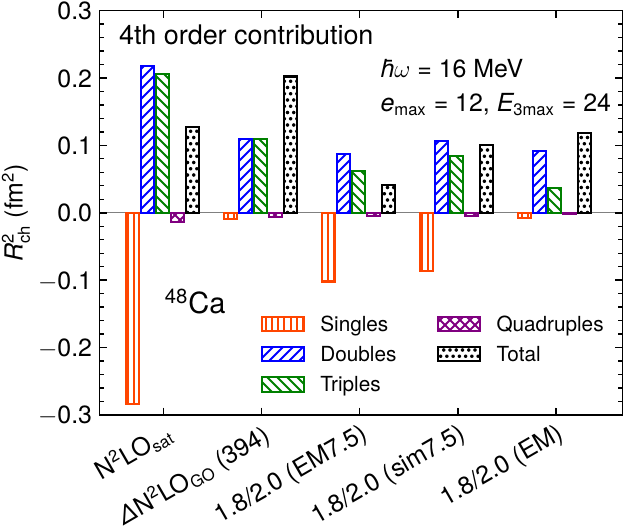} 
\caption{Contributions from singles, doubles, triples, and quadruples to the fourth-order correction of the charge radius squared of $^{48}$Ca for different interactions.}
\label{fig:o4_all_int}
\end{figure}

We start by investigating different correlations included in the fourth-order correction of MBPT. To this end, we characterize the fourth-order diagrams by their $n$-particle--$n$-hole ($npnh$) excitation rank. This yields a set of diagrams containing $1p1h$ (singles), $2p2h$ (doubles), $3p3h$ (triples), and $4p4h$ (quadruples) excitations between the second and third vertices of the diagrams. A full list of fourth-order diagrams can be found in Fig.~19 of Ref.~\cite{Hergert2016_IMSRG}. The $1p1h$ and $2p2h$ intermediate excitations are also present in the second and third orders, while the $3p3h$ and $4p4h$ excitations enter only at fourth order and beyond. As illustrated in Fig.~\ref{fig:o4_all_nuclei}, singles, doubles, and triples correlations produce negative energy corrections, whereas quadruples yield positive energy corrections, across different nuclei using the \magicint{} interaction. While triples and quadruples are the dominant contributions, their effects largely cancel, resulting in a small net contribution. Similar trends are observed for the other ``softer'' interactions, as shown in Fig.~\ref{fig:e4_all_int}. A major cancellation occurs among triples and quadruples resulting in a net attractive correction which is close to zero for the case of the \deltago{} interaction. The situation changes qualitatively for the ``harder'' \nnlosat{} interaction. While the cancellation persists, the net fourth-order contribution becomes positive in this case. We emphasize that our findings are consistent with the general understanding from non-perturbative CC calculations, which typically require the inclusion of triples corrections to accurately describe ground-state energies of atomic nuclei~\cite{Hagen2014}. Since the CCSD approximation includes all fourth-order MBPT diagrams except those involving triples excitations, our analysis shows that the leading triples corrections provide an additional attractive contribution to the total ground-state energy. This is consistent with results obtained in CC theory~\cite{Hagen2014,Binder2014,Novario2020a}.

We also analyze the fourth-order correction to the charge radius of $^{48}$Ca. The decomposition is depicted in Fig.~\ref{fig:o4_all_int}, which now also includes the non-canonical MBPT diagrams due to the presence of a non-trivial one-body vertex from the radius operator. We observe that the quadruples generate very small negative contributions for all interactions. The contribution from singles varies for the different interactions. The doubles and triples yield a similar positive correction, and their total effect dominates the fourth-order contribution, resulting in a positive net correction from the fourth-order diagrams.

\subsection{Comparison with IMSRG(2)}
\label{sec:mbpt_vs_imsrg2}

The non-perturbative IMSRG method resums large classes of MBPT diagrams to all orders (for details see~\cite{Hergert2016_IMSRG}). Specifically, the IMSRG(2) truncation retains effective operators up to the NO2B level throughout the flow, incorporating all MBPT diagrams up to third order. In addition, it includes part of the fourth-order MBPT diagrams (namely, those involving the single-, double- and selected quadruple-excitation contributions), as well as the particle-particle ($pp$) and hole-hole ($hh$) ladder diagrams, the particle-hole ring diagrams (hereafter referred to as $ph$ ladder diagrams), the $pp/hh$-$ph$ interference diagrams and some additional ring diagrams that can be constructed by commuting two-body operators. As shown in Ref.~\cite{Hergert2016_IMSRG}, the triples appearing at the fourth order are missing in the IMSRG(2) approximation, while the quadruples come with a slight mismatch in the prefactor, $Q_a+Q_b/2$, thus undercounting the diagram class $Q_b$, where $Q_a+Q_b$ consists of the full set of quadruples at the fourth order with 
\begin{eqnarray}
\vcenter{\hbox{\includegraphics[height=2.75cm]{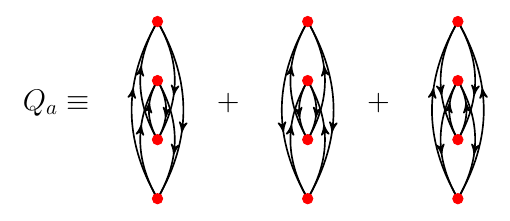}}},
\end{eqnarray}
and 
\begin{eqnarray}
\vcenter{\hspace{-4mm}\hbox{\includegraphics[height=2.75cm]{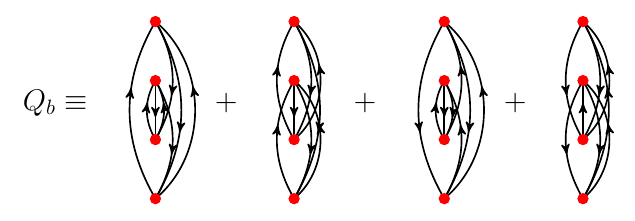}}}\hspace{-8mm}.\hspace{8mm}
\end{eqnarray}

\begin{figure}[t!]
\includegraphics[width=0.45\textwidth]{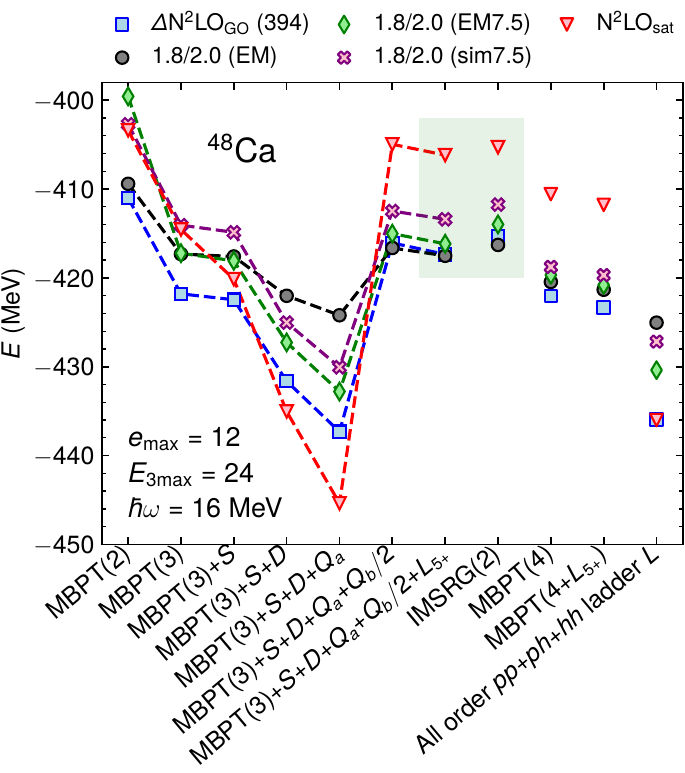}
\caption{MBPT results with different groups of diagrams included for the ground-state energy of $^{48}$Ca using different interactions in comparison to IMSRG(2). $S$, $D$, $Q$, and $L_{5+}$ are singles, doubles, quadruples in the fourth-order correction, and the sum of ladder diagrams from fifth to infinite order, respectively.}
\label{fig:mbpt_vs_imsrg2}
\end{figure}

A detailed comparison between the MBPT and IMSRG(2) calculations of the $^{48}$Ca ground-state energy for the five different Hamiltonians is given in Fig.~\ref{fig:mbpt_vs_imsrg2}. As can be seen from the figure, the MBPT(3) calculation may deviate substantially from the IMSRG(2) result, especially for the ``harder'' \nnlosat{} interaction. In addition, MBPT(3) predicts a different sequence of ground-state energies across the Hamiltonians than IMSRG(2). By gradually including the fourth-order singles ($S$), doubles ($D$), and quadruples $Q_a$ on top of the MBPT(3) results, the ground-state energy decreases for all interactions. 
However, the  $Q_b$ diagrams contribute positively to the ground-state energies (see Fig.~\ref{fig:mbpt_vs_imsrg2}), while as noted above, IMSRG(2) includes only half of $Q_b$. This may partially compensate for the omission of the attractive triples contributions and thus prevent an underestimation of the ground-state energy. In contrast, the CCSD truncation includes all fourth-order contributions except those involving triple excitations, resulting thus in a higher ground-state energy than that from IMSRG(2). This can be seen in Fig.~\ref{fig:Ca48_with_different_int} for the case of the \magicint{} interaction.

After adding all the $pp$, $ph$, and $hh$ ladder diagrams resummed from fifth to infinite order  (denoted as $L_{5+}$), the difference between MBPT and IMSRG(2) is within about 2\,MeV for each interaction, including the \nnlosat{} interaction, as indicated by the shaded area. In addition, the ordering of ground-state energies for the different interactions remains the same. The 2\,MeV difference mentioned above is expected to arise from additional $pp/hh$-$ph$ interference diagrams and ring diagrams included in IMSRG(2). The further inclusion of both triples and $Q_b/2$, corresponding thus to MBPT(4+$L_{5+}$), leads to more negative ground-state energies and inverts the ordering of those obtained from the \magicint{} and \deltago{} interactions.

\begin{figure}[t!]
\includegraphics[width=0.425\textwidth]{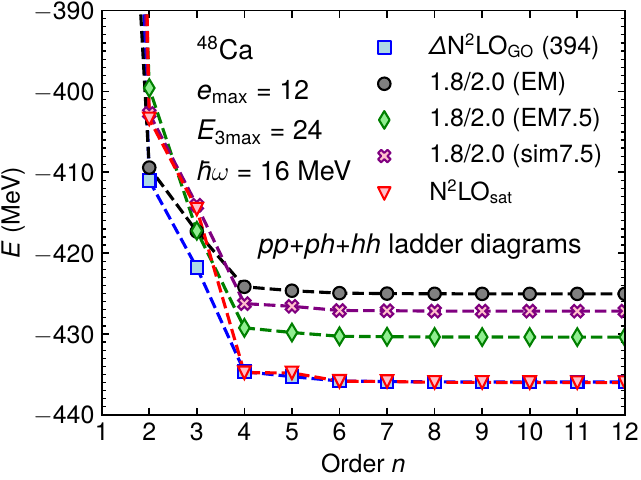}
\caption{Sum of $pp$, $ph$, and $hh$ ladder diagrams as a function of perturbative order $n$ for the ground-state energy of $^{48}$Ca calculated for different interactions.}
\label{fig:ladder}
\end{figure}

The discussed ladder diagram results are obtained through a perturbative resummation of ladder diagrams until convergence is reached. Their numerical evaluation is straightforward, as higher-order ladder diagrams can be efficiently evaluated through repeated matrix-matrix (or matrix-vector) products. Figure~\ref{fig:ladder} shows the sum of $pp$, $ph$, and $hh$ ladder diagrams as a function of perturbative order, where the (first-order) HF energy is included. Therefore, all diagrams up to the third order are included in this summation. As shown in the figure, the sum of ladder diagrams converges quickly in terms of perturbative order. The converged results yield significantly more negative ground-state energies than MBPT(4) and IMSRG(2) as shown in the last column of Fig.~\ref{fig:mbpt_vs_imsrg2}. 

The comparison with the non-perturbative IMSRG(2) results highlights that the MBPT calculations in Fig.~\ref{fig:mbpt_vs_imsrg2} are reliable (at least up to the fourth order) for the interactions considered in this work. As a result, nuclear ground states can be computed with higher precision  within MBPT(4) [and possibly MBPT(5*)], compared to the conventional MBPT(3) approximation. Furthermore, the difference between MBPT(3)+$S$+$D$+$Q_a$+$Q_b/2$+$L_{5+}$ and MBPT(4+$L_{5+}$) suggests that the triples and partial quadruples omitted in IMSRG(2) provide an additional 4--6\,MeV binding to $^{48}$Ca, depending on the interaction. 

\section{Conclusion and outlook}
\label{sec:summary}

In this work, we have advanced MBPT for atomic nuclei to fifth order and achieved higher-precision calculations of nuclear ground states by implementing automated diagram and code generation. A multilevel truncation scheme that employs an order-dependent model-space truncation is proposed and applied to calculations of the ground-state energy and nuclear radii for closed-shell nuclei up to $^{78}$Ni. A clear convergence trend in terms of perturbative order is observed for the ground-state energy, while a somewhat slower convergence is obtained for the radius. These findings are robust for a variety of ``softer'' chiral NN+3N interactions.
We also note that, for the radius, the interaction uncertainty is likely larger than the uncertainty associated with the perturbative truncation. For the ground-state energy, the magnitude of the fourth-order correction is generally less than half of the third-order one, but the ratio to the third-order correction is larger than that between the third- and second-order corrections. The detailed decomposition of the fourth-order corrections reveals that the attractive triples and repulsive quadruples are the dominant correlations in the fourth-order energy correction, and that their effects strongly cancel. Our high-order MBPT calculations enable a detailed analysis of the diagrammatic content of the IMSRG(2) approximation and the relative importance of particular classes of diagrams for the total value of the observable. This allows to estimate the missing correlations at the IMSRG(2) level due to fourth-order triples and quadruples. A statistical uncertainty quantification of MBPT, as explored in Ref.~\cite{Svensson2026}, with the input from the fourth- and fifth-order calculation will be interesting. In addition, the MBPT derivation of valence-space effective Hamiltonian can also be extended beyond third order following the same strategy~\cite{Li2023thesis}. We leave these developments to future work.

\begin{acknowledgements}
We thank F.~Bonaiti and M.~Heinz for providing CC and IMSRG results and M.~Cincar, T.~Duguet, M.~Heinz, F.~Marino, T.~Miyagi, U.~Vernik, J.~Vary, and L.~Zurek for helpful discussions. The nuclear interactions were generated with the \texttt{NuHamil} code~\cite{Miyagi2023}, and the IMSRG calculations were performed with the \texttt{IMSRG++} code~\cite{imsrgpp}. This work was supported in part by the European Research Council (ERC) under the European Union's Horizon 2020 research and innovation programme (Grant Agreement No.~101020842) and under the European Union's Horizon Europe research and innovation programme (Grant Agreement No.~101162059), and by CNRS/IN2P3, France, via ABI-CONFI Master project. The authors gratefully acknowledge the Gauss Centre for Supercomputing e.V. (www.gauss-centre.eu) for providing computing time through the John von Neumann Institute for Computing (NIC) on the GCS Supercomputer JUWELS at J\"ulich Supercomputing Centre (JSC), as well as computing time provided on the high-performance computers Lichtenberg II at TU Darmstadt and M\'esocentre de Calcul Intensif Aquitain (MCIA) at University of Bordeaux. 
\end{acknowledgements}

\bibliography{ref}

@STRING(PLB="Phys. Lett. B")

@STRING(EPJA="Eur. Phys. J. A")

@STRING(PR="Phys. Rep.")

@STRING(PPNP="Prog. Part. Nucl. Phys.")

@STRING(nucl-th="LANL archive nucl-th")

@STRING(CPCommun="Comput. Phys. Commun.")

@STRING(APNY="Ann. Phys. (N.Y.)")

@article{Miyagi:2023zvv,
    author = "Miyagi, T. and Cao, X. and Seutin, R. and Bacca, S. and Garcia Ruiz, R. F. and Hebeler, K. and Holt, J. D. and Schwenk, A.",
    title = "{Impact of Two-Body Currents on Magnetic Dipole Moments of Nuclei}",
    doi = "10.1103/PhysRevLett.132.232503",
    journal = "Phys. Rev. Lett.",
    volume = "132",
    number = "23",
    pages = "232503",
    year = "2024"
}

@article{Door:2024qqz,
    author = "Door, Menno and Yeh, C. H. and Heinz, M. and Kirk, F. and Lyu, C. and Miyagi, T. and Berengut, J. C. and Biero{\'n}, J. and Blaum K. and L.~S.~Dreissen, L. S. and others",
    title = "{Probing New Bosons and Nuclear Structure with Ytterbium Isotope Shifts}",
    doi = "10.1103/PhysRevLett.134.063002",
    journal = "Phys. Rev. Lett.",
    volume = "134",
    number = "6",
    pages = "063002",
    year = "2025"
}

@misc{Kuske:2025tsm,
      title={$r$-process nucleosynthesis with \textit{ab initio} nuclear masses around the {$N=82$} shell closure}, 
      author={Jan Kuske and Takayuki Miyagi and Almudena Arcones and Achim Schwenk},
      year={2025},
      eprint={2509.19131},
      archivePrefix={arXiv},
      primaryClass={astro-ph.HE}
}

@article{Keller:2022crb,
    author = "Keller, J. and Hebeler, K. and Schwenk, A.",
    title = "{Nuclear Equation of State for Arbitrary Proton Fraction and Temperature Based on Chiral Effective Field Theory and a Gaussian Process Emulator}",
    doi = "10.1103/PhysRevLett.130.072701",
    journal = "Phys. Rev. Lett.",
    volume = "130",
    number = "7",
    pages = "072701",
    year = "2023"
}

@article{Alp:2025wjn,
    author = "Alp, Faruk and Dietz, Yannick and Hebeler, Kai and Schwenk, Achim",
    title = "{Equation~of state and Fermi liquid properties of dense matter based on chiral effective field theory interactions}",
    doi = "10.1103/ls3l-dn1y",
    journal = "Phys. Rev. C",
    volume = "112",
    number = "5",
    pages = "055802",
    year = "2025"
}

@misc{Arthuis2024Hamil,
    author = "Arthuis, P. and Hebeler, K. and Schwenk, A.",
    title = "{Neutron-rich nuclei and neutron skins from chiral low-resolution interactions}",
    eprint = "2401.06675",
    archivePrefix = "arXiv",
    primaryClass = "nucl-th",
    year = "2024"
}

@article{Paldus1973,
title = "{Computer generation of Feynman diagrams for perturbation theory {I}. General algorithm}",
journal = {Comput. Phys. Commun.},
volume = {6},
number = {1},
pages = {1-7},
year = {1973},
issn = {0010-4655},
doi = {https://doi.org/10.1016/0010-4655(73)90016-7},
url = {https://www.sciencedirect.com/science/article/pii/0010465573900167},
author = {J. Paldus and H. C. Wong},
}

@article{Janssen1991,
	author = {Janssen, Curtis L. and Schaefer, Henry F.},
	date = {1991/01/01},
	doi = {10.1007/BF01113327},
	id = {Janssen1991},
	isbn = {1432-2234},
	journal = {Theor. Chim. Acta},
	number = {1},
	pages = {1--42},
	title = {The automated solution of second quantization equations with applications to the coupled cluster approach},
	url = {https://doi.org/10.1007/BF01113327},
	volume = {79},
	year = {1991}
}

@article{Li1994,
    author = {Li, Xiangzhu and Paldus, Josef},
    title = {Automation of the implementation of spin‐adapted open‐shell coupled‐cluster theories relying on the unitary group formalism},
    journal = {J. Chem. Phys.},
    volume = {101},
    number = {10},
    pages = {8812-8826},
    year = {1994},
    month = {11},
    issn = {0021-9606},
    doi = {10.1063/1.468074}
}

@article{Novario2020a,
  title = {Charge radii of exotic neon and magnesium isotopes},
  author = {Novario, S. J. and Hagen, G. and Jansen, G. R. and Papenbrock, T.},
  journal = {Phys. Rev. C},
  volume = {102},
  issue = {5},
  pages = {051303(R)},
  numpages = {8},
  year = {2020},
  month = {Nov},
  publisher = {American Physical Society},
  doi = {10.1103/PhysRevC.102.051303},
  url- = {https://link.aps.org/doi/10.1103/PhysRevC.102.051303}
}

@article{Marino2026,
  title = {From closed shells to open shells: Coupled-cluster calculations of atomic nuclei},
  author = {Marino, F. and Bonaiti, F. and Demol, P. and Bacca, S. and Duguet, T. and Hagen, G. and Jansen, G. R. and Papenbrock, T. and Tichai, A.},
  journal = {Phys. Rev. C},
  volume = {113},
  issue = {4},
  pages = {044301},
  numpages = {13},
  year = {2026},
  month = {Apr},
  publisher = {American Physical Society},
  doi = {10.1103/p297-y8vq},
  url = {https://link.aps.org/doi/10.1103/p297-y8vq}
}

@misc{Chen2026qcombo,
      title={Qcombo: A Python Package for Automated Commutator Calculations of Quantum Many-Body Operators}, 
      author={L. H. Chen and Y. Li and H. Hergert and J. M. Yao},
      year={2026},
      eprint={2603.24399},
      archivePrefix={arXiv},
      primaryClass={nucl-th}
}

@article{Evangelista2022,
    author = {Evangelista, Francesco A.},
    title = {Automatic derivation of many-body theories based on general {Fermi} vacua},
    journal = {J. Chem. Phys.},
    volume = {157},
    number = {6},
    pages = {064111},
    year = {2022},
    month = {08},
    issn = {0021-9606},
    doi = {10.1063/5.0097858}
}

@misc{Bonaiti2025Pb,
    author = "Bonaiti, Francesca and Hagen, Gaute and Papenbrock, Thomas",
    title = "{Structure of the doubly magic nuclei $^{208}$Pb and $^{266}$Pb from \textit{ab initio} computations}",
    eprint = "2508.14217",
    archivePrefix = "arXiv",
    primaryClass = "nucl-th",
    month = "8",
    year = {2025}
}

@misc{Drischler2026,
    author = "Drischler, C. and McElvain, K. S. and Arthuis, P.",
    title = "{Many-body perturbation theory for the nuclear equation of state up to fifth order}",
    eprint = "2603.24532",
    archivePrefix = "arXiv",
    primaryClass = "nucl-th",
    year = {2026}
}

@article{Hoferichter:2020osn,
    author = "Hoferichter, Martin and Men{\'e}ndez, Javier and Schwenk, Achim",
    title = "{Coherent elastic neutrino-nucleus scattering: EFT analysis and nuclear responses}",
    doi = "10.1103/PhysRevD.102.074018",
    journal = "Phys. Rev. D",
    volume = "102",
    number = "7",
    pages = "074018",
    year = "2020"
}

@article{imsrgpp, 
author = {Stroberg, S. R.}, 
journal = "\url{https://github.com/ragnarstroberg/imsrg}"
}

@article{Svensson2026,
  title = {Bayesian approach for many-body uncertainties in nuclear structure: Many-body perturbation theory for finite nuclei},
  author = {Svensson, I. and Tichai, A. and Hebeler, K. and Schwenk, A.},
  journal = {Phys. Rev. C},
  volume = {113},
  issue = {2},
  pages = {024303},
  numpages = {12},
  year = {2026},
  month = {Feb},
  publisher = {American Physical Society},
  doi = {10.1103/y8kt-mgf5},
  url = {https://link.aps.org/doi/10.1103/y8kt-mgf5}
}

@article{Kuo1981,
title = "{A simple method for evaluating Goldstone diagrams in an angular momentum coupled representation}",
journal = APNY,
volume = {132},
number = {2},
pages = {237-276},
year = {1981},
issn = {0003-4916},
doi = {https://doi.org/10.1016/0003-4916(81)90068-3},
url = {https://www.sciencedirect.com/science/article/pii/0003491681900683},
author = {T.T.S Kuo and J Shurpin and K.C Tam and E Osnes and P.J Ellis}
}

@article{Friar1997,
  title = {Nuclear sizes and the isotope shift},
  author = {Friar, J. L. and Martorell, J. and Sprung, D. W. L.},
  journal = {Phys. Rev. A},
  volume = {56},
  issue = {6},
  pages = {4579--4586},
  numpages = {0},
  year = {1997},
  month = {Dec},
  publisher = {American Physical Society},
  doi = {10.1103/PhysRevA.56.4579},
  url = {https://link.aps.org/doi/10.1103/PhysRevA.56.4579}
}

@article{Dickhoff2004,
title = {Self-consistent {Green}'s function method for nuclei and nuclear matter},
journal = PPNP,
volume = {52},
number = {2},
pages = {377-496},
year = {2004},
issn = {0146-6410},
doi = {https://doi.org/10.1016/j.ppnp.2004.02.038},
url = {https://www.sciencedirect.com/science/article/pii/S0146641004000535},
author = {W.H. Dickhoff and C. Barbieri}
}

@article{Hergert2016_IMSRG,
title = {The {In-Medium Similarity Renormalization Group}: A novel {\it ab initio} method for nuclei},
journal = PR,
volume = {621},
pages = {165-222},
year = {2016},
issn = {0370-1573},
doi = {https://doi.org/10.1016/j.physrep.2015.12.007},
url = {https://www.sciencedirect.com/science/article/pii/S0370157315005414},
author = {H. Hergert and S. K. Bogner and T. D. Morris and A. Schwenk and K. Tsukiyama}
}

@ARTICLE{Hergert2020_review,
AUTHOR={Hergert, Heiko},   
TITLE={A Guided Tour of {\it ab initio} Nuclear Many-Body Theory},      
JOURNAL={Front. Phys.},      
VOLUME={8}, 
PAGES={379},
YEAR={2020}, 
DOI={10.3389/fphy.2020.00379},
URL={https://www.frontiersin.org/articles/10.3389/fphy.2020.00379},             
ISSN={2296-424X}
}

@article{Hu2022ab,
  title="{Ab initio predictions link the neutron skin of $^{208}$Pb to nuclear forces}",
  author={Hu, Baishan and Jiang, Weiguang and Miyagi, Takayuki and Sun, Zhonghao and Ekstr{\"o}m, Andreas and Forss{\'e}n, Christian and Hagen, Gaute and Holt, Jason D and Papenbrock, Thomas and Stroberg, S Ragnar and others},
  journal={Nat. Phys.},
  volume={18},
  number={10},
  pages={1196--1200},
  year={2022},
  doi={https://doi.org/10.1038/s41567-022-01715-8}, 
  url={10.1038/s41567-022-01715-8}, 
  publisher={Nature Publishing Group UK London},
  note="[Erratum: \href{https://doi.org/10.1038/s41567-023-02324-9}{Nat. Phys. 20, 169 (2024)}]"
}

@article{Jiang2020,
  title = "{Accurate bulk properties of nuclei from $A=2$ to $\ensuremath{\infty}$ from potentials with $\mathrm{\ensuremath{\Delta}}$ isobars}",
  author = {Jiang, W. G. and Ekstr\"om, A. and Forss\'en, C. and Hagen, G. and Jansen, G. R. and Papenbrock, T.},
  journal = {Phys. Rev. C},
  volume = {102},
  issue = {5},
  pages = {054301},
  numpages = {8},
  year = {2020},
  month = {Nov},
  publisher = {American Physical Society},
  doi = {10.1103/PhysRevC.102.054301},
  url = {https://link.aps.org/doi/10.1103/PhysRevC.102.054301}
}

@article{Hebeler2023,
  title = {Normal ordering of three-nucleon interactions for \textit{ab initio} calculations of heavy nuclei},
  author = {Hebeler, K. and Durant, V. and Hoppe, J. and Heinz, M. and Schwenk, A. and Simonis, J. and Tichai, A.},
  journal = {Phys. Rev. C},
  volume = {107},
  issue = {2},
  pages = {024310},
  numpages = {14},
  year = {2023},
  month = {Feb},
  publisher = {American Physical Society},
  doi = {10.1103/PhysRevC.107.024310},
  url = {https://link.aps.org/doi/10.1103/PhysRevC.107.024310}
}

@BOOK{Shavitt,
   AUTHOR="I. Shavitt and R.J. Bartlett",
   TITLE="Many-Body Methods in Chemistry and Physics: MBPT and Coupled-Cluster Theory",
   PUBLISHER="Cambridge University Press",
   ADDRESS="Cambridge",
   YEAR=2009
}

@phdthesis{Li2023thesis,
    author = "Li, Zhen",
    title = "{Advances in many-body perturbation theory for closed- and open-shell nuclei}",
    school = "University of Bordeaux",
    year = {2023},
    url={https://theses.hal.science/tel-04813436v1}
}

@article{Tichai2020AMC,
    author = "Tichai, A. and Wirth, R. and Ripoche, J. and Duguet, T.",
    title = "{Symmetry reduction of tensor networks in many-body theory I. Automated symbolic evaluation of SU(2) algebra}",
    doi = "10.1140/epja/s10050-020-00233-6",
    journal = "Eur. Phys. J. A",
    volume = "56",
    number = "10",
    pages = "272",
    year = "2020"
}

@ARTICLE{Goldstone1957,
   AUTHOR="J. Goldstone",
   TITLE="{Derivation of the Brueckner many-body theory}", 
   URL="https://doi.org/10.1098/rspa.1957.0037", 
   DOI="10.1098/rspa.1957.0037", 
   JOURNAL="Proc. Roy. Soc. A",
   VOLUME=239,
   PAGES=267,
   YEAR=1957
}

@article{Brandow1967,
  title = {Linked-Cluster Expansions for the Nuclear Many-Body Problem},
  author = {Brandow, Baird H.},
  journal = {Rev. Mod. Phys.},
  volume = {39},
  issue = {4},
  pages = {771--828},
  numpages = {0},
  year = {1967},
  month = {Oct},
  publisher = {American Physical Society},
  doi = {10.1103/RevModPhys.39.771},
  url = {https://link.aps.org/doi/10.1103/RevModPhys.39.771}
}

@article{Tichai2018,
title = {Bogoliubov many-body perturbation theory for open-shell nuclei},
journal = PLB,
volume = {786},
pages = {195-200},
year = {2018},
issn = {0370-2693},
doi = {https://doi.org/10.1016/j.physletb.2018.09.044},
url = {https://www.sciencedirect.com/science/article/pii/S0370269318307457},
author = {A. Tichai and P. Arthuis and T. Duguet and H. Hergert and V. Somà and R. Roth}
}

@article{Tichai2023epe,
  title="{Towards heavy-mass {\it ab initio} nuclear structure: Open-shell {Ca, Ni and Sn} isotopes from Bogoliubov coupled-cluster theory}",
journal = {Phys. Lett. B},
volume = {851},
pages = {138571},
year = {2024},
doi = {https://doi.org/10.1016/j.physletb.2024.138571},
author = {A. Tichai and P. Demol and T. Duguet}
}

@article{Soma2011,
  title = {{\it Ab initio} self-consistent {Gorkov-Green}'s function calculations of semimagic nuclei: Formalism at second order with a two-nucleon interaction},
  author = {Som\`a, V. and Duguet, T. and Barbieri, C.},
  journal = {Phys. Rev. C},
  volume = {84},
  issue = {6},
  pages = {064317},
  numpages = {33},
  year = {2011},
  month = {Dec},
  publisher = {American Physical Society},
  doi = {10.1103/PhysRevC.84.064317},
  url = {https://link.aps.org/doi/10.1103/PhysRevC.84.064317}
}

@article{Barbieri2022,
  title = {Gorkov algebraic diagrammatic construction formalism at third order},
  author = {Barbieri, Carlo and Duguet, Thomas and Som\`a, Vittorio},
  journal = {Phys. Rev. C},
  volume = {105},
  issue = {4},
  pages = {044330},
  numpages = {29},
  year = {2022},
  month = {Apr},
  publisher = {American Physical Society},
  doi = {10.1103/PhysRevC.105.044330},
  url = {https://link.aps.org/doi/10.1103/PhysRevC.105.044330}
}

@article{Hebeler2011_EM1.8_2.0,
  title = {Improved nuclear matter calculations from chiral low-momentum interactions},
  author = {Hebeler, K. and Bogner, S. K. and Furnstahl, R. J. and Nogga, A. and Schwenk, A.},
  journal = {Phys. Rev. C},
  volume = {83},
  issue = {3},
  pages = {031301},
  numpages = {5},
  year = {2011},
  month = {Mar},
  publisher = {American Physical Society},
  doi = {10.1103/PhysRevC.83.031301},
  url = {https://link.aps.org/doi/10.1103/PhysRevC.83.031301}
}

@article{Langhammer2012_Pade,
  title = "{Spectra of open-shell nuclei with Pad\'e-resummed degenerate perturbation theory}",
  author = {Langhammer, Joachim and Roth, Robert and Stumpf, Christina},
  journal = {Phys. Rev. C},
  volume = {86},
  issue = {5},
  pages = {054315},
  numpages = {8},
  year = {2012},
  month = {Nov},
  publisher = {American Physical Society},
  doi = {10.1103/PhysRevC.86.054315},
  url = {https://link.aps.org/doi/10.1103/PhysRevC.86.054315}
}

@article{Drischler2019,
  title = {Chiral Interactions up to Next-to-Next-to-Next-to-Leading Order and Nuclear Saturation},
  author = {Drischler, C. and Hebeler, K. and Schwenk, A.},
  journal = {Phys. Rev. Lett.},
  volume = {122},
  issue = {4},
  pages = {042501},
  numpages = {6},
  year = {2019},
  month = {Jan},
  publisher = {American Physical Society},
  doi = {10.1103/PhysRevLett.122.042501},
  url = {https://link.aps.org/doi/10.1103/PhysRevLett.122.042501}
}

@article{Arthuis2019ADG1,
title = "{{ADG}: Automated generation and evaluation of many-body diagrams I. Bogoliubov many-body perturbation theory}",
journal = CPCommun,
volume = {240},
pages = {202-227},
year = {2019},
issn = {0010-4655},
doi = {https://doi.org/10.1016/j.cpc.2018.11.023},
url = {https://www.sciencedirect.com/science/article/pii/S0010465518304156},
author = {P. Arthuis and T. Duguet and A. Tichai and R.-D. Lasseri and J.-P. Ebran}
}

@article{Brolli2025,
  title = "{Diagrammatic Monte Carlo for Finite Systems at Zero Temperature}",
  author = {Brolli, Stefano and Barbieri, Carlo and Vigezzi, Enrico},
  journal = {Phys. Rev. Lett.},
  volume = {134},
  issue = {18},
  pages = {182502},
  numpages = {6},
  year = {2025},
  month = {May},
  publisher = {American Physical Society},
  doi = {10.1103/PhysRevLett.134.182502}
}

@article{Zhen2026,
  title = {Stochastic many-body perturbation theory for high-order calculations},
  author = {Zhen, X. and Hu, R. Z. and Pei, J. C. and Xu, F. R.},
  journal = {Phys. Rev. C},
  volume = {113},
  issue = {5},
  pages = {L051302},
  numpages = {7},
  year = {2026},
  month = {May},
  publisher = {American Physical Society},
  doi = {10.1103/q3vn-8y8s}
}

@article{Demol2026,
title = "{\textit{Ab initio} calculations of nuclear charge radii across and beyond $^{132}$Sn: Putting chiral EFT nuclear interactions to the test}",
journal = {Phys. Lett. B},
volume = {878},
pages = {140524},
year = {2026},
issn = {0370-2693},
doi = {https://doi.org/10.1016/j.physletb.2026.140524},
url = {https://www.sciencedirect.com/science/article/pii/S0370269326003771},
author = {P. Demol and U. Vernik and T. Duguet and A. Tichai}
}

@article{Arthuis2021ADG2,
title = "{{ADG}: Automated generation and evaluation of many-body diagrams {II}. Particle-number projected Bogoliubov many-body perturbation theory}",
journal = CPCommun,
volume = {261},
pages = {107677},
year = {2021},
issn = {0010-4655},
doi = {https://doi.org/10.1016/j.cpc.2020.107677},
url = {https://www.sciencedirect.com/science/article/pii/S0010465520303295},
author = {P. Arthuis and A. Tichai and J. Ripoche and T. Duguet}
}

@article{Tichai2022ADG3,
  title="{{ADG}: automated generation and evaluation of many-body diagrams: {III}. Bogoliubov in-medium similarity renormalization group formalism}",
  author={Tichai, A and Arthuis, P and Hergert, H and Duguet, T},
  journal=EPJA,
  volume={58},
  number={1},
  pages={2},
  year={2022},
  url={https://doi.org/10.1140/epja/s10050-021-00621-6}, 
  doi={10.1140/epja/s10050-021-00621-6}, 
  publisher={Springer}
}

@article{Binder2014,
title = {{\it Ab initio} path to heavy nuclei},
journal = PLB,
volume = {736},
pages = {119-123},
year = {2014},
issn = {0370-2693},
doi = {https://doi.org/10.1016/j.physletb.2014.07.010},
url = {https://www.sciencedirect.com/science/article/pii/S0370269314004961},
author = {Sven Binder and Joachim Langhammer and Angelo Calci and Robert Roth}
}

@article{Roth2010,
title = {Pad\'e-resummed high-order perturbation theory for nuclear structure calculations},
journal = PLB,
volume = {683},
number = {4},
pages = {272-277},
year = {2010},
issn = {0370-2693},
doi = {https://doi.org/10.1016/j.physletb.2009.12.046},
url = {https://www.sciencedirect.com/science/article/pii/S037026930901507X},
author = {Robert Roth and Joachim Langhammer}
}

@article{Tichai2016,
title = "{Hartree-Fock many-body perturbation theory for nuclear ground-states}",
journal = PLB,
volume = {756},
pages = {283-288},
year = {2016},
issn = {0370-2693},
doi = {https://doi.org/10.1016/j.physletb.2016.03.029},
url = {https://www.sciencedirect.com/science/article/pii/S0370269316002008},
author = {Alexander Tichai and Joachim Langhammer and Sven Binder and Robert Roth}
}

@article{Hu2016,
  title = "{{\it Ab initio} nuclear many-body perturbation calculations in the Hartree-Fock basis}",
  author = {Hu, B. S. and Xu, F. R. and Sun, Z. H. and Vary, J. P. and Li, T.},
  journal = {Phys. Rev. C},
  volume = {94},
  pages = {014303},
  year = {2016},
  doi = {10.1103/PhysRevC.94.014303},
  url = {https://link.aps.org/doi/10.1103/PhysRevC.94.014303}
}

@ARTICLE{Tichai2020,
AUTHOR={Tichai, Alexander and Roth, Robert and Duguet, Thomas},
TITLE={Many-Body Perturbation Theories for Finite Nuclei},
JOURNAL={Front. Phys.},
VOLUME={8},
pages = {164}, 
YEAR={2020},
URL={https://www.frontiersin.org/articles/10.3389/fphy.2020.00164},
DOI={10.3389/fphy.2020.00164},
ISSN={2296-424X}
}

@article{Hagen2014,
doi = {10.1088/0034-4885/77/9/096302},
url = {https://dx.doi.org/10.1088/0034-4885/77/9/096302},
year = {2014},
month = {sep},
publisher = {IOP Publishing},
volume = {77},
number = {9},
pages = {096302},
author = {G Hagen and T Papenbrock and M Hjorth-Jensen and D J Dean},
title = {Coupled-cluster computations of atomic nuclei},
journal = {Rep. Prog. Phys.}
}

@article{Machleidt2011,
title = {Chiral effective field theory and nuclear forces},
journal = PR,
volume = {503},
number = {1},
pages = {1-75},
year = {2011},
issn = {0370-1573},
doi = {https://doi.org/10.1016/j.physrep.2011.02.001},
url = {https://www.sciencedirect.com/science/article/pii/S0370157311000457},
author = {R. Machleidt and D.R. Entem}
}

@article{Epelbaum2009,
  title = {Modern theory of nuclear forces},
  author = {Epelbaum, E. and Hammer, H.-W. and Mei\ss{}ner, Ulf-G.},
  journal = {Rev. Mod. Phys.},
  volume = {81},
  issue = {4},
  pages = {1773--1825},
  numpages = {0},
  year = {2009},
  month = {Dec},
  publisher = {American Physical Society},
  doi = {10.1103/RevModPhys.81.1773},
  url = {https://link.aps.org/doi/10.1103/RevModPhys.81.1773}
}

@article{Hebeler2021,
title = {Three-nucleon forces: Implementation and applications to atomic nuclei and dense matter},
journal = PR,
volume = {890},
pages = {1-116},
year = {2021},
doi = {https://doi.org/10.1016/j.physrep.2020.08.009},
url = {https://www.sciencedirect.com/science/article/pii/S0370157320303409},
author = {Kai Hebeler}
}

@article{Simonis2019,
    author = "Simonis, Johannes and Bacca, Sonia and Hagen, Gaute",
    title = "{First principles electromagnetic responses in medium-mass nuclei}",
    doi = "10.1140/epja/i2019-12825-0",
    journal = "Eur. Phys. J. A",
    volume = "55",
    number = "12",
    pages = "241",
    year = "2019"
}

@article{BonaitiCCresults,
author="F. Bonaiti",
title="",
doi="",
journal="",
volume="",
pages="",
year="",
note="{private communication (2026)}"
}

@article{Hagen2015Ca48,
    author = "Hagen, G. and Ekstr{\"o}m, A. and Forss{\'en}, C. and Jansen, G. R. and Nazarewicz, W. and Papenbrock, T. and Wendt, K. A. and Bacca, S. and Barnea, N. and Carlsson, B. and others",
    title = "{Neutron and weak-charge distributions of the $^{48}$Ca nucleus}",
    doi = "10.1038/nphys3529",
    journal = "Nat. Phys.",
    volume = "12",
    number = "2",
    pages = "186",
    year = "2015"
}

@article{Suzuki1980,
    author = {Suzuki, Kenji and Lee, Shyh Yuan},
    title = {Convergent Theory for Effective Interaction in Nuclei},
    journal = {Prog. Theor. Phys.},
    volume = {64},
    number = {6},
    pages = {2091-2106},
    year = {1980},
    month = {12},
    issn = {0033-068X},
    doi = {10.1143/PTP.64.2091}
}

@article{Brandow1977,
title = {Linked-Cluster Perturbation Theory for Closed- and Open-Shell Systems},
journal = {Adv. Quantum Chem.},
volume = {10},
pages = {187-249},
year = {1977},
doi = {https://doi.org/10.1016/S0065-3276(08)60581-X},
author = {B.H. Brandow}
}

@article{Lietz2016,
    author = "Lietz, Justin and Novario, Sam and Jansen, Gustav R. and Hagen, Gaute and Hjorth-Jensen, Morten",
    title = {Computational Nuclear Physics and Post {Hartree-Fock} Methods},
    doi = "10.1007/978-3-319-53336-0_8",
    journal = "Lect. Notes Phys.",
    volume = "936",
    pages = "293--399",
    year = "2017"
}

@article{Ekstrom2015_NNLOsat,
  title = {Accurate nuclear radii and binding energies from a chiral interaction},
  author = {Ekstr\"om, A. and Jansen, G. R. and Wendt, K. A. and Hagen, G. and Papenbrock, T. and Carlsson, B. D. and Forss\'en, C. and Hjorth-Jensen, M. and Navr\'atil, P. and Nazarewicz, W.},
  journal = {Phys. Rev. C},
  volume = {91},
  issue = {5},
  pages = {051301},
  numpages = {7},
  year = {2015},
  month = {May},
  publisher = {American Physical Society},
  doi = {10.1103/PhysRevC.91.051301},
  url = {https://link.aps.org/doi/10.1103/PhysRevC.91.051301},
  note = "[Erratum: \href{https://doi.org/10.1103/PhysRevC.109.059901}{Phys. Rev. C 109, 059901 (2024)}]"
}

@article{Furnstahl2012,
  title = {Corrections to nuclear energies and radii in finite oscillator spaces},
  author = {Furnstahl, R. J. and Hagen, G. and Papenbrock, T.},
  journal = {Phys. Rev. C},
  volume = {86},
  issue = {3},
  pages = {031301},
  numpages = {5},
  year = {2012},
  month = {Sep},
  publisher = {American Physical Society},
  doi = {10.1103/PhysRevC.86.031301},
  url = {https://link.aps.org/doi/10.1103/PhysRevC.86.031301}
}

@article{Barrett2013,
title = {{\it Ab initio} no core shell model},
journal = PPNP,
volume = {69},
pages = {131-181},
year = {2013},
issn = {0146-6410},
doi = {https://doi.org/10.1016/j.ppnp.2012.10.003},
url = {https://www.sciencedirect.com/science/article/pii/S0146641012001184},
author = {Bruce R. Barrett and Petr Navr\'atil and James P. Vary}
}

@article{Stroberg2019_review_Heff,
author = {Stroberg, S. Ragnar and Hergert, Heiko and Bogner, Scott K. and Holt, Jason D.},
title = {Nonempirical Interactions for the Nuclear Shell Model: An Update},
journal = {Annu. Rev. Nucl. Part. Sci.},
volume = {69},
number = {1},
pages = {307-362},
year = {2019},
doi = {10.1146/annurev-nucl-101917-021120},
URL = {https://doi.org/10.1146/annurev-nucl-101917-021120}
}

@article{Bloch1958,
	title = {Sur la th{\'e}orie des perturbations des {\'e}tats li{\'e}s},
	volume = {6},
	issn = {00295582},
	url = {https://linkinghub.elsevier.com/retrieve/pii/0029558258901160},
	doi = {10.1016/0029-5582(58)90116-0},
	journal = {Nucl. Phys.},
	author = {Bloch, Claude},
	month = mar,
	year = {1958},
	pages = {329--347}
}

@article{MHJ1995,
title = {Realistic effective interactions for nuclear systems},
journal = PR,
volume = {261},
number = {3},
pages = {125-270},
year = {1995},
issn = {0370-1573},
doi = {https://doi.org/10.1016/0370-1573(95)00012-6},
url = {https://www.sciencedirect.com/science/article/pii/0370157395000126},
author = {Morten Hjorth-Jensen and Thomas T.S. Kuo and Eivind Osnes} 
}

@article{Coraggio2009,
title = {Shell-model calculations and realistic effective interactions},
journal = PPNP,
volume = {62},
number = {1},
pages = {135-182},
year = {2009},
issn = {0146-6410},
doi = {https://doi.org/10.1016/j.ppnp.2008.06.001},
url = {https://www.sciencedirect.com/science/article/pii/S0146641008000410},
author = {L. Coraggio and A. Covello and A. Gargano and N. Itaco and T.T.S. Kuo}
}

@article{Bogner2014,
  title = "{Nonperturbative Shell-Model Interactions from the {In-Medium Similarity Renormalization Group}}",
  author = {Bogner, S. K. and Hergert, H. and Holt, J. D. and Schwenk, A. and Binder, S. and Calci, A. and Langhammer, J. and Roth, R.},
  journal = {Phys. Rev. Lett.},
  volume = {113},
  issue = {14},
  pages = {142501},
  numpages = {5},
  year = {2014},
  month = {Oct},
  publisher = {American Physical Society},
  doi = {10.1103/PhysRevLett.113.142501},
  url = {https://link.aps.org/doi/10.1103/PhysRevLett.113.142501}
}

@article{Ong2010,
  title = {Effect of spin-orbit nuclear charge density corrections due to the anomalous magnetic moment on halonuclei},
  author = {Ong, A. and Berengut, J. C. and Flambaum, V. V.},
  journal = {Phys. Rev. C},
  volume = {82},
  issue = {1},
  pages = {014320},
  numpages = {6},
  year = {2010},
  month = {Jul},
  publisher = {American Physical Society},
  doi = {10.1103/PhysRevC.82.014320},
  url = {https://link.aps.org/doi/10.1103/PhysRevC.82.014320}
}

@article{PDG2024,
    author = "Navas, S. and others",
    collaboration = "Particle Data Group",
    title = "{Review of particle physics}",
    doi = "10.1103/PhysRevD.110.030001",
    journal = "Phys. Rev. D",
    volume = "110",
    number = "3",
    pages = "030001",
    year = "2024"
}

@article{Tsukiyama2011,
  title = {In-Medium Similarity Renormalization Group For Nuclei},
  author = {Tsukiyama, K. and Bogner, S. K. and Schwenk, A.},
  journal = {Phys. Rev. Lett.},
  volume = {106},
  issue = {22},
  pages = {222502},
  numpages = {4},
  year = {2011},
  month = {Jun},
  publisher = {American Physical Society},
  doi = {10.1103/PhysRevLett.106.222502},
  url = {https://link.aps.org/doi/10.1103/PhysRevLett.106.222502}
}

@article{Miyagi2023,
    author = "Miyagi, Takayuki",
    title = "{NuHamil: A numerical code to generate nuclear two- and three-body matrix elements from chiral effective field theory}",
    doi = "10.1140/epja/s10050-023-01039-y",
    journal = "Eur. Phys. J. A",
    volume = "59",
    number = "7",
    pages = "150",
    year = "2023"
}

@article{Miyagi2022_E3max,
  title = {Converged {\it ab initio} calculations of heavy nuclei},
  author = {Miyagi, T. and Stroberg, S. R. and Navr\'atil, P. and Hebeler, K. and Holt, J. D.},
  journal = {Phys. Rev. C},
  volume = {105},
  issue = {1},
  pages = {014302},
  numpages = {14},
  year = {2022},
  month = {Jan},
  publisher = {American Physical Society},
  doi = {10.1103/PhysRevC.105.014302},
  url = {https://link.aps.org/doi/10.1103/PhysRevC.105.014302}
}

@article{Heinz2021,
  title = {In-medium similarity renormalization group with three-body operators},
  author = {Heinz, M. and Tichai, A. and Hoppe, J. and Hebeler, K. and Schwenk, A.},
  journal = {Phys. Rev. C},
  volume = {103},
  issue = {4},
  pages = {044318},
  numpages = {21},
  year = {2021},
  month = {Apr},
  publisher = {American Physical Society},
  doi = {10.1103/PhysRevC.103.044318},
  url = {https://link.aps.org/doi/10.1103/PhysRevC.103.044318}
}

@article{Heinz2025,
  title = {Improved structure of calcium isotopes from {\it ab initio} calculations},
  author = {Heinz, M. and Miyagi, T. and Stroberg, S. R. and Tichai, A. and Hebeler, K. and Schwenk, A.},
  journal = {Phys. Rev. C},
  volume = {111},
  issue = {3},
  pages = {034311},
  numpages = {14},
  year = {2025},
  month = {Mar},
  publisher = {American Physical Society},
  doi = {10.1103/PhysRevC.111.034311},
  url = {https://link.aps.org/doi/10.1103/PhysRevC.111.034311}
}

@article{He2024,
  title = "{Factorized approximation to the in-medium similarity renormalization group IMSRG(3)}",
  author = {He, B. C. and Stroberg, S. R.},
  journal = {Phys. Rev. C},
  volume = {110},
  issue = {4},
  pages = {044317},
  numpages = {16},
  year = {2024},
  month = {Oct},
  publisher = {American Physical Society},
  doi = {10.1103/PhysRevC.110.044317},
  url = {https://link.aps.org/doi/10.1103/PhysRevC.110.044317}
}

@article{Stroberg2024,
  title = {In-medium similarity renormalization group with flowing 3-body operators, and approximations thereof},
  author = {Stroberg, S. R. and Morris, T. D. and He, B. C.},
  journal = {Phys. Rev. C},
  volume = {110},
  issue = {4},
  pages = {044316},
  numpages = {16},
  year = {2024},
  month = {Oct},
  publisher = {American Physical Society},
  doi = {10.1103/PhysRevC.110.044316},
  url = {https://link.aps.org/doi/10.1103/PhysRevC.110.044316}
}

@article{Simonis:2017dny,
    author = "Simonis, J. and Stroberg, S. R. and Hebeler, K. and Holt, J. D. and Schwenk, A.",
    title = "{Saturation with chiral interactions and consequences for finite nuclei}",
    doi = "10.1103/PhysRevC.96.014303",
    journal = "Phys. Rev. C",
    volume = "96",
    number = "1",
    pages = "014303",
    year = "2017"
}

\end{document}